\documentclass[12pt]{article}
\usepackage{titling}
\usepackage{amsmath}
\usepackage{slashed}
\usepackage{amssymb}
\usepackage{epsfig}
\usepackage{graphicx}
\usepackage{multirow}
\usepackage{color}
\usepackage{subcaption}
\usepackage{rotating}
\usepackage{hyperref}
\usepackage[margin=1.0in]{geometry}
\usepackage[table]{xcolor}
\usepackage{enumitem}
\usepackage[utf8x]{inputenc}
\usepackage[compress,numbers,sort]{natbib}
\usepackage{authblk}
\usepackage{colortbl}
\usepackage{pdflscape}
\usepackage{color}

\definecolor{nicered}{rgb}{0.6,0.1,0.1}
\definecolor{nicegreen}{rgb}{0.1,0.5,0.1}
\definecolor{mediumcandyapplered}{rgb}{0.99, 0.12, 0.07}
\definecolor{red}{rgb}{1.0, 0, 0}
\hypersetup{colorlinks,citecolor= nicegreen,linkcolor= nicered}

\renewcommand{\bar}{\overline}

\definecolor{LightCyan}{rgb}{0.88,1,1}
\definecolor{piggypink}{rgb}{0.99, 0.87, 0.9}
\definecolor{applegreen}{rgb}{0.55, 0.71, 0.0}
\definecolor{darkpastelgreen}{rgb}{0.01, 0.75, 0.24}
\definecolor{green-yellow}{rgb}{0.68, 1.0, 0.18}

\newcommand{\beq}{\begin{equation}}
\newcommand{\eeq}{\end{equation}}
\newcommand{\bea}{\begin{eqnarray}}
\newcommand{\eea}{\end{eqnarray}}

\title{\bf{Light-quark Yukawa couplings from off-shell Higgs production}}

\author[]
{Elisa Balzani \thanks{elisa.balzani@pd.infn.it}}
\author[]
{Ramona Gr\"ober \thanks{ramona.groeber@pd.infn.it}}
\author[]
{Marco Vitti \thanks{marco.vitti@pd.infn.it}}

\affil[]{\emph{\normalsize Dipartimento di Fisica e Astronomia ``G. Galilei'', Universit\`a di Padova, Italy, and Istituto Nazionale di Fisica Nucleare, Sezione di Padova, I-35131 Padova, Italy.}}

\date{}

\begin{document}

\maketitle
\begin{abstract}
\normalsize
Yukawa couplings of the first quark generation are notoriously difficult to constrain due to their small values within the Standard Model.
Here we propose Higgs off-shell production, with the Higgs boson decaying to four leptons,
 as a probe of the up- and down-quark Yukawa couplings. 
Using kinematic discriminants similar to the ones employed in the Higgs width measurements 
we find that the down (up) Yukawa coupling can be constrained to a factor of 156 (260) times its Standard Model value at the
high-luminosity LHC assuming only experimental systematic uncertainties. 
Off-shell Higgs production hence provides better sensitivity to the first-generation quark Yukawa couplings
with respect to other probes such as Higgs+jet or Higgs pair production.
\end{abstract}

\clearpage


\clearpage

\section{Introduction}
Ten years after the Higgs boson discovery, several challenges must still be addressed in order to fully understand the properties of this particle. While the Higgs boson couplings to gauge bosons and to third-generation fermions  have been measured at the $\mathcal{O}(5 -20\%)$ level \cite{ATLAS:2022vkf, CMS:2022dwd},  little is known about the Higgs boson couplings to first- and second-generation quarks and leptons, with the exception of the Higgs coupling to muons \cite{ATLAS:2020fzp, CMS:2020xwi}. Even at the high-luminosity LHC (HL-LHC) the first- and second-generation quark Yukawa couplings remain rather elusive: in a global fit it was found that the modification factor $\kappa_q=y_q/y_q^\text{SM}$ of the quark Yukawa coupling $y_q$ with respect to its Standard Model (SM) value $y_q^\text{SM}$ can be constrained to $\kappa_u<560$, $\kappa_d<260$, $\kappa_s<13$ and $\kappa_c<1.2$ \cite{deBlas:2019rxi}. The fit is not entirely model-independent but assumes that the light-quark Yukawa couplings can be constrained from the Higgs untagged branching ratio. 
\par
On the other hand, there exist various proposals on how to constrain the light Yukawa couplings from measurements of specific processes. For instance, the Higgs boson coupling to the charm quark has been constrained to be smaller than 8.5 times its SM value \cite{ATLAS:2022ers, CMS:2019hve} by looking at associated $Vh$ production\footnote{$Vh$ production followed by the Higgs decay to two jets has been studied as a probe of light Yukawa couplings in ref.~\cite{Carpenter:2016mwd}.} with subsequent decay of the Higgs boson to charm quarks~\cite{Perez:2015lra}. 
Also exclusive Higgs decays to vector mesons \cite{Bodwin:2013gca, Kagan:2014ila, Konig:2015qat, Alte:2016yuw} have been used to constrain the charm Yukawa coupling \cite{Aaboud:2018txb, CMS:2022fsq}. Other proposals include Higgs+charm production \cite{Brivio:2015fxa}, 
the change in the Higgs $p_T$-spectrum from enhanced charm-quark loops \cite{Bishara:2016jga} or  $VVcj$ production \cite{Vignaroli:2022fqh}. 
Constraining the strange Yukawa coupling is extremely challenging, but at a future $e^+e^-$ collider one might reach SM sensitivity if strange tagging is employed \cite{Duarte-Campderros:2018ouv}.  
\par
These ideas are mostly based on the Higgs decays. For the first quark generation one needs to proceed in a different manner since Higgs decays to light quarks cannot be measured directly.  One can though make use of the 
fact that for enhanced light-quark Yukawa couplings a significant contribution of  Higgs production can come from diagrams where the Higgs boson couples directly to the quark content of the parton distributions. 
Interesting processes that can be used to constrain the light-quark Yukawa couplings in this way are Higgs+photon \cite{Aguilar-Saavedra:2020rgo}, Higgs+jet \cite{Soreq:2016rae, Bonner:2016sdg}, Higgs pair production \cite{Alasfar:2019pmn, Alasfar:2022vqw}, tri-vector boson production in the high-energy limit \cite{Falkowski:2020znk, Vignaroli:2022fqh} and the charge asymmetry in $W^{\pm}h$ \cite{Yu:2017vul, Yu:2016rvv}.
\par 
In this paper we want to study another probe of the first-generation quark Yukawa couplings, namely the measurement of an off-shell Higgs boson decaying to a $Z$ boson pair that subsequently decays to leptons. A study of light-quark Yukawa couplings for the $h\to ZZ$ final state  has been presented in ref.~\cite{Zhou:2015wra} for the 7 and 8 TeV runs of the LHC. In our analysis we reappraise the idea of ref.~\cite{Zhou:2015wra} in light of the evidence for off-shell production found by recent measurements \cite{CMS:2022ley, ATLAS:2023dnm} and show that the use of kinematic discriminants can significantly improve the projected limits on $\kappa_q$ at the HL-LHC.   
\par
The off-shell Higgs measurement is usually considered in combination with an on-shell measurement as a probe of the Higgs total width \cite{Kauer:2012hd, Caola:2013yja, Campbell:2013una}, for which a direct determination is not possible at the LHC. In particular, the Higgs width can be indirectly constrained by the ratio of on- and off-shell signal strengths
\begin{equation}
\frac{\mu_\text{on}}{\mu_\text{off}}\propto\frac{\kappa_{ggh}^2(m_h) \kappa_{hZZ}^2(m_h)}{\Gamma_h/\Gamma_{h}^\text{SM}}\frac{1}{\kappa_{ggh}^2(m_{4\ell}) \kappa_{hZZ}^2(m_{4\ell})}\,, \label{eq:onoff}
\end{equation}
where $\kappa_i$ with $i=ggh, hZZ$ are the coupling modifiers with respect to the SM value and $\Gamma_h$ is the Higgs width.
The invariant mass of the four-lepton pair is denoted by $m_{4\ell}$, the Higgs mass by $m_h$.
Under the limitation that the effective coupling of the Higgs to gluons and  that the coupling of the Higgs to $Z$ bosons have a predictable energy dependence \cite{Englert:2014aca,Englert:2014ffa},  the Higgs width can be extracted. 
The CMS collaboration has recently measured $\Gamma_h$ to be $3.2^{+2.4}_{-1.7}\text{ MeV}$  \cite{CMS:2022ley} employing this method; ATLAS obtained $4.5^{+3.3}_{-2.5} \text{ MeV}$ \cite{ATLAS:2023dnm}. Both measurements are in agreement with the SM value.
\par 
In the presence of enhanced light-quark Yukawa couplings, though, the above method of constraining the Higgs width cannot be applied straightforwardly. While enhanced light-quark Yukawa couplings increase the Higgs total width, in addition the aforementioned production of a Higgs boson directly from light-quark fusion needs to be considered. 
Adding this extra production channel falsifies the assumption on which the interpretation of eq.~\eqref{eq:onoff} in terms of a measurement of the Higgs width is based. 
 On the other hand, the new production channel has different kinematic properties with respect to the dominant off-shell Higgs production channel in the SM, namely gluon fusion. This observation can be used to set bounds on the light-quark Yukawa couplings.
\par
This paper is structured as follows: in section \ref{sec:lightYuk} we  introduce our setup of enhanced light-quark Yukawa couplings and in particular we show how the latter can be modified by new physics effects while still having small SM quark masses. In section \ref{sec:onshell} we  discuss the on-shell Higgs measurement in this context. In section \ref{sec:offshell}
we describe our calculation of the off-shell Higgs cross sections, before we present in section \ref{sec:analysis} the results of our analysis. We conclude in section \ref{sec:con}. 
 

\section{Light Yukawa couplings in Effective Field Theory \label{sec:lightYuk}}
The Higgs couplings to quarks in the SM are described by the Lagrangian
\begin{equation}
\mathcal{L}_y=-y_{i j}^u \bar{Q}_L^i \tilde{\phi} u_R^j-y_{i j}^d \bar{Q}_L^i \phi d_R^j+\text { h.c. }\,,
\end{equation} 
where $\tilde{\phi}=i \sigma_2 \phi^*$, $\sigma_2$ is the second Pauli matrix, $\phi$ represents the Higgs doublet, $Q_L^i$ the left-handed $SU(2)$  quark doublet of the $i$-th generation and $u_R^j$ and $d_R^j$ the right-handed up- and down-type fields of the $j$-th generation, respectively. Deformations of the SM in a model-independent way, under the assumption that the Higgs field transforms as a SM doublet, can be studied in the Standard Model Effective Field Theory (SMEFT), parametrizing the effects of new physics (NP) with higher-dimensional operators suppressed by some large energy scale $\Lambda$. A complete basis of higher-dimensional operators has been given in refs.~\cite{Grzadkowski:2010es, Contino:2013kra}. In this work we are interested only in an enhancement of the light-quark Yukawa couplings. We hence do not consider operators that require a redefinition of the Higgs field and that lead to a general shift of the Higgs couplings, as they cannot render the light-quark Yukawa couplings sufficiently larger and at the same time obey the limits on other Higgs couplings. 

In the SMEFT, new flavour structures can be introduced through dimension-six operators that contain flavour indices. Especially, the couplings of the quarks are modified by the operator
\begin{equation}
\Delta \mathcal{L}_y=\frac{\phi^{\dagger} \phi}{\Lambda^2}\left((C_{u\phi})_{i j} \bar{Q}_L^i \tilde{\phi} u_R^j+(C_{d\phi})_{i j} \bar{Q}_L^i \phi d_R^j+\text { h.c. }\right), 
\end{equation} 
with $i,j=1,...,3$ and $\Lambda$ denoting the cut-off of the effective field theory (EFT). Here $u$ and $d$ refer to the up- and down-type sectors respectively, and not the quarks themselves. Therefore the mass matrices of the up-type and down-type quarks obtained from the Yukawa and the new SMEFT couplings are
\begin{equation}
\begin{aligned}
&M_{i j}^u=\frac{v}{\sqrt{2}}\left(y_{i j}^u-\frac{1}{2} (C_{u\phi})_{i j} \frac{v^2}{\Lambda^2}\right), \\
&M_{i j}^d=\frac{v}{\sqrt{2}}\left(y_{i j}^d-\frac{1}{2} (C_{d\phi})_{i j} \frac{v^2}{\Lambda^2}\right) .
\end{aligned}
\end{equation}
Due to the modification of the mass matrix, the rotation matrices transforming quark wavefunctions to the mass eigenbasis will be modified with respect to the SM ones. Mass matrices are diagonalized by a new set of bi-unitary transformations 
\begin{equation}
m_{q_i}=\left(\left(V_L^{u/d}\right)^\dagger M^{u/d}V_R^{u/d}\right)_{ii},
\end{equation}  
in which the CKM matrix is defined as $V_\text{CKM}=\left(V_L^u\right)^\dagger V_L^d$. We can rewrite $ (C_{q\phi})_{i j}$ in terms of $(\tilde{C}_{q\phi})_{ij}$ which are now in the mass eigenbasis
\begin{equation}
(\tilde{C}_{q\phi})_{ij}=\left(V_L^q\right)_{n i}^* (C_{q\phi})_{n m} \left(V_R^q\right)_{m j}, \ \ \ \ \ \text{with} \ \ \ q=u, d\,.
\end{equation}
Therefore the Lagrangian containing the couplings of the Higgs boson to quarks is
\begin{equation}
\mathcal{L}\supset g_{hq_{i}\bar{q}_{j}} \bar{q}_j q_i h+ g_{hhq_{i}\bar{q}_{j}}      \bar{q}_j q_i h^2 + g_{hhhq_{i}\bar{q}_{j}}      \bar{q}_j q_i h^3
\end{equation}
with
\begin{equation}
g_{hq_{i}\bar{q}_{j}}= \frac{m_q}{v}\delta_{ij}-\frac{1}{\sqrt{2}}\frac{v^2}{\Lambda^2}{(\tilde{C}_{q\phi})_{ij}}\,, \hspace*{0.5cm}g_{hhq_{i}\bar{q}_{j}}=-\frac{3}{2\sqrt{2}}\frac{v}{\Lambda^2}{(\tilde{C}_{q\phi})_{ij}}\,, \hspace*{0.5cm}g_{hhhq_{i}\bar{q}_{j}}=- \frac{1}{2\sqrt{2}}\frac{1}{\Lambda^2}{(\tilde{C}_{q\phi})_{ij}}\, , \label{eq:ghqq}
\end{equation}
where $i, j=  1, \dots, 3$ are generation indices. An important feature of this parametrisation is that the NP effects encoded in $(\tilde{C}_{q\phi})_{ij}$ are kept separate from the contribution due to the quark mass in the first of eqs.~\eqref{eq:ghqq}. Therefore, the coupling of the Higgs boson to quarks can receive large enhancements even if the quark kinetic masses are small.
For later use, we give also the coupling of quarks to two neutral Goldstone bosons
\begin{equation}
g_{G_0 G_0q_{i}\bar{q}_{j}}= -\frac{1}{2\sqrt{2}}\frac{v}{\Lambda^2}{(\tilde{C}_{q\phi})_{ij}}\,. 
\end{equation}
As we will see later, employing the Goldstone boson equivalence theorem for the off-shell Higgs measurement we can observe a similar behaviour as in Higgs pair production \cite{Alasfar:2019pmn, Alasfar:2022vqw}, which motivates the study of light Yukawa couplings in this context.
\par
Finally, we introduce here another convenient notation for the coupling of the Higgs boson to quarks, resembling  the ``kappa"-framework (see e.g.~\cite{deBlas:2019rxi}). This notation is valid only for the diagonal couplings and does not allow for new Lorentz structures with respect to the SM ones: 
\begin{equation}
g_{hq\bar{q}}=\kappa_q \frac{m_q}{v}\,. \label{eq:ghqqkappa}
\end{equation}
We define the reference mass values $m_u=2.2\text{ MeV}$ and $m_d=4.7\text{ MeV}$ as constant,  i.e.~not running, values. 
In the rest of the paper, we will often use $\kappa_q$ values when quoting sensitivity limits, as this is often more intuitive with respect to the theoretically well-defined SMEFT approach. The given values need  though always to be understood with respect to the reference mass values. We remark that the limits on $\kappa_q$ values can be directly translated to limits on the SMEFT coefficients $\tilde{C}_{q\phi}/\Lambda^2$ ($q=u,d$) using eq.~\eqref{eq:ghqq} and eq.~\eqref{eq:ghqqkappa}, and that the two notations that we have introduced are equivalent.
\par
We finally shortly comment on models that can achieve large deviations in the light-quark Yukawa couplings. Large deviations can be implemented using an aligned flavour assumption, as new sources of flavour violation are constrained to the level $|(\tilde{C}_{u\phi/d\phi})_{12}|<10^{-5}\Lambda^2/v^2$  $|(\tilde{C}_{u\phi/d\phi})_{13}|<10^{-4}\Lambda^2/v^2$ by $\Delta F=2$ transitions \cite{Blankenburg:2012ex, Harnik:2012pb}. From an EFT perspective there is no general reason why an aligned flavour should not be implemented, as the renormalisation group flow introduces flavour-changing couplings that are suppressed by several mass insertions and hence strongly suppressed for the first generation \cite{Jenkins:2013wua}. It is hence rather a question of a concrete ultraviolet model that implements the aligned flavour assumption with large deviations in the light-quark Yukawa couplings at the electroweak scale. In a simplified model approach, an effective $|\phi|^2 \bar{Q}_L \phi d_R$ or $|\phi|^2 \bar{Q}_L \tilde{\phi} u_R$ operator can be generated by various fields, e.g.~vector-like fermions, new scalars and new vector bosons and combinations thereof. Concrete models have been presented in refs.~\cite{Egana-Ugrinovic:2018znw, Egana-Ugrinovic:2019dqu, Bar-Shalom:2018rjs}. We will in the following refrain from using concrete models but note that those models will at the same time generate other effective operators, which possibly can be constrained  better than the light-quark Yukawa couplings. 

\section{The on-shell Higgs \label{sec:onshell}}
As stated in the introduction, an indirect determination of the total decay width of the Higgs boson can be extracted from a comparison of off-shell and on-shell measurements. In this section we shortly discuss the effects of enhanced light Yukawa couplings in the case of on-shell Higgs production, focusing on the subsequent decay of the Higgs to a $Z$ pair.  We omit the treatment of the leptonic decays of the $Z$ bosons, since these are not affected by the NP contributions that we are studying. 
In the context of on-shell production, we use the narrow width approximation to express the cross section as the product of the Higgs cross section and the $h\to ZZ$ branching ratio
\begin{equation}
\sigma(pp \to h \to ZZ) = \sigma(pp \to h) \cdot \text{BR}_{ZZ},
\label{eq:nwa}
\end{equation} 
and we consider the effects of modified light-quark Yukawa couplings in each contribution of the r.h.s.~of eq.~\eqref{eq:nwa}.

\begin{figure}[t]
\centering
\begin{subfigure}{.5\textwidth}
  \centering
  \includegraphics[scale=0.8]{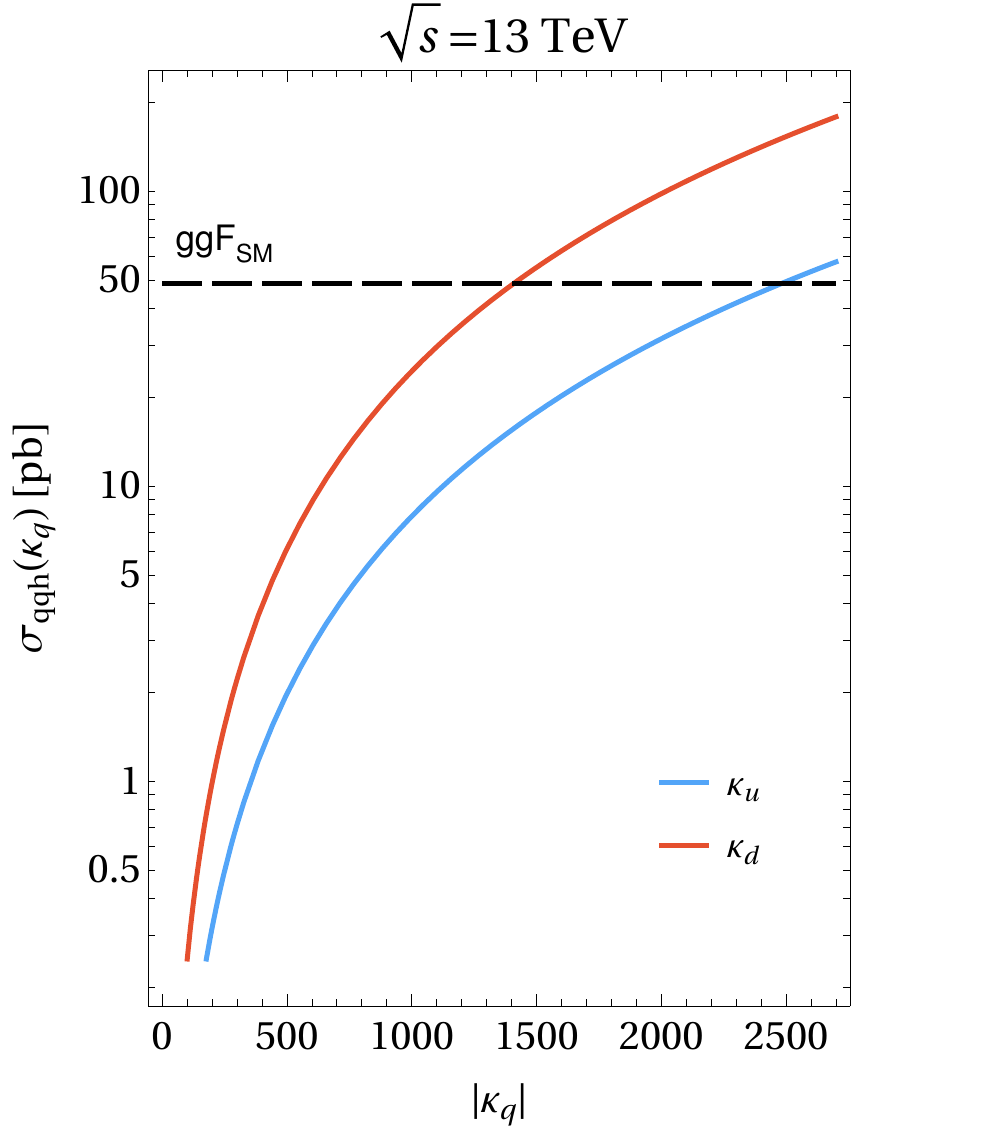} 
  \caption{}
  \label{fig:qqhos}
\end{subfigure}%
\begin{subfigure}{.5\textwidth}
  \centering
  \includegraphics[scale=0.82]{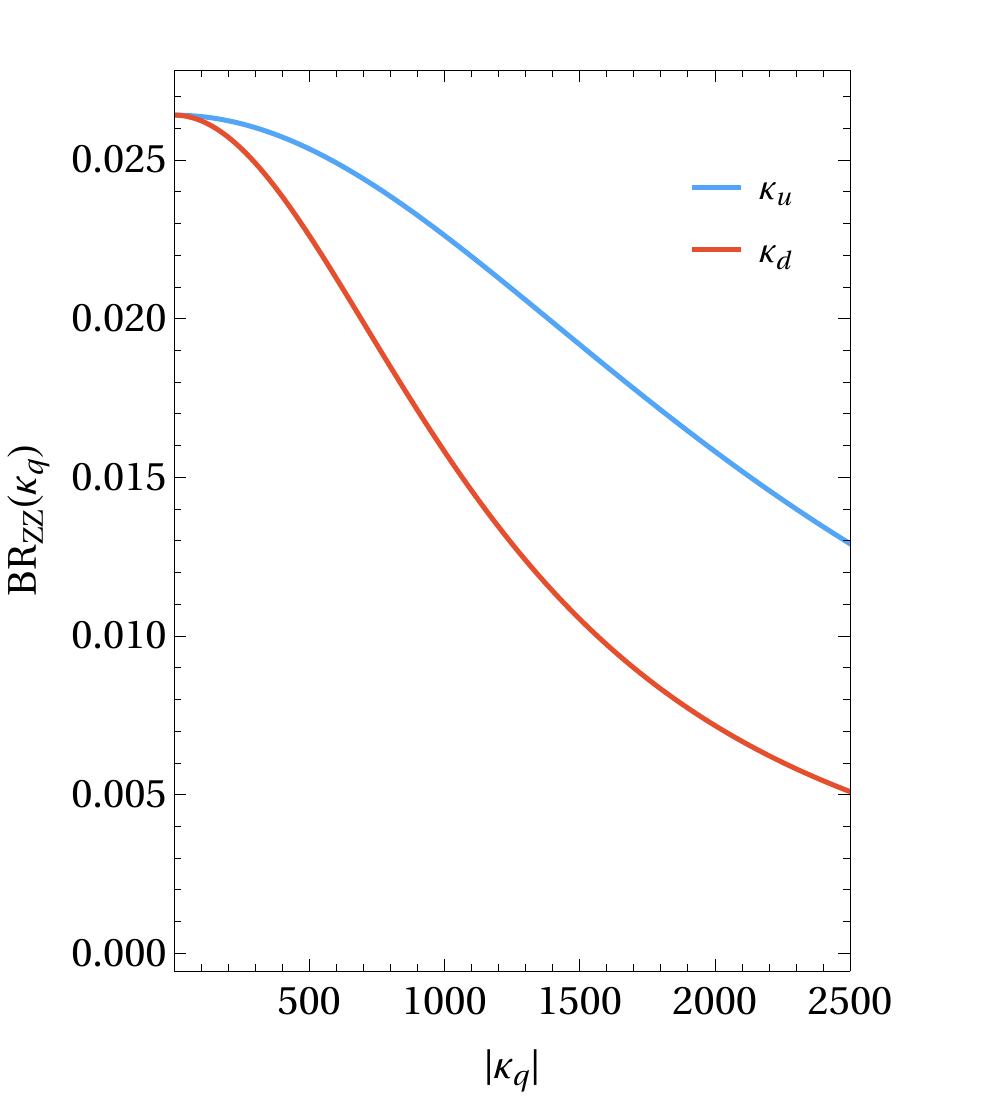} 
  \caption{}
  \label{fig:br}
\end{subfigure}\\
\caption{(a) Modification of the cross section for resonant $q \bar{q} \to h$ production, as a function of the light-Yukawa coupling modifier $\kappa_q$. The cross section for the gluon-fusion channel in the SM \cite{Cepeda:2019klc} is shown as a black dashed line. (b) Branching ratio for the $h \to ZZ$ decay as a function of $\kappa_q$ for the first quark generation.}
\label{fig:qqh_br}
\end{figure}

Whereas in the SM the only important partonic channel for Higgs production at the LHC is the gluon fusion (ggF) channel via quarks of the third generation (the top/bottom interference amounts to about 10\%), if we allow the light-quark Yukawa couplings to be larger than their SM value we can expect new  relevant contributions via loops of first- and second-generation quarks. In addition, a new partonic channel must be considered for Higgs production, namely $q \bar{q}$ annihilation, $q \bar{q}\to h$. In our study we computed the ggF cross section at leading order (LO) and we used a $K$-factor to account for the N$^3$LO QCD corrections~\cite{Cepeda:2019klc, Anastasiou:2016cez} by rescaling the LO result.   We observed that the effects of modified first-generation Yukawa couplings, stemming from the top/light-quark interference, are below the level of 0.1\% for $|\kappa_d| \lesssim 1200$ and $|\kappa_u| \lesssim 4500$, therefore we consider them negligible\footnote{The modification of the strange Yukawa coupling can reach an effect of 3\% on the ggF cross section for $|\kappa_s|=200$, while an enhancement of the charm Yukawa coupling by $|\kappa_c|=20$ leads to a change of about 18\%.}, both in this section and in the following one.

On the other hand, the $q \bar{q} \to h$ channel becomes increasingly 
more important for modified first-generation Yukawa couplings. In this case, the quadratic enhancement with $\kappa_q$ of the partonic cross section can effectively compensate the suppression of the quark luminosities with respect to ggF, and from fig.~\ref{fig:qqhos} 
one can notice that $q \bar{q} \to h$ becomes the dominant Higgs production mode at the 13 TeV LHC for values of $\kappa_d$ ($\kappa_u$) larger than about 1400 (2500). Similarly to ggF production, we used a $K$-factor to rescale the LO cross section in order to account for the NLO QCD corrections. The value $K_{q\bar{q}h}^\text{NLO}= 1.4$ has been obtained by adapting the calculation of the QCD corrections of $b \bar{b} \to h$ in refs.~\cite{Dicus:1998hs, Balazs:1998sb, Harlander:2003ai}.

Finally, the last piece to be considered in our on-shell prediction is the modified $ZZ$ branching ratio, $\text{BR}_{ZZ}(\kappa_q)$. We used $\texttt{HDECAY}$ \cite{Djouadi:1997yw, Djouadi:2018xqq} to verify that the impact of large values of $\kappa_q$ on the partial width $\Gamma(h \to ZZ)$ via higher-order corrections is negligible. At the same time, the total decay width, $\Gamma_h^\text{BSM} (\kappa_q)$, is substantially increased in the case of enhanced $\kappa_q$ because the Higgs decay channels to the first-generation quarks become relevant. Their partial width can be simply expressed by rescaling the SM one by a factor $\kappa_q^2$. We then have
\begin{equation}
\Gamma_h^\text{BSM} (\kappa_q) = \Gamma_h^\text{SM} + \kappa_q^2 ~\Gamma^\text{SM}(h \to q \bar{q}) \qquad (q=u, d),
\label{eq:gamma_bsm}
\end{equation}
with $\Gamma_h^\text{SM}= 4.1~\text{MeV}$ \cite{LHCHiggsCrossSectionWorkingGroup:2016ypw}. 
As a consequence of eq.~\eqref{eq:gamma_bsm}, the $h \to ZZ$ branching ratio is reduced with growing $\kappa_q$. In fig.~\ref{fig:br} we show the modifications of $\text{BR}_{ZZ}(\kappa_q)$ due to the separate enhancement of the down (red) and up (blue) Yukawa couplings.

\begin{figure}[t]
\centering
\includegraphics[scale=0.8]{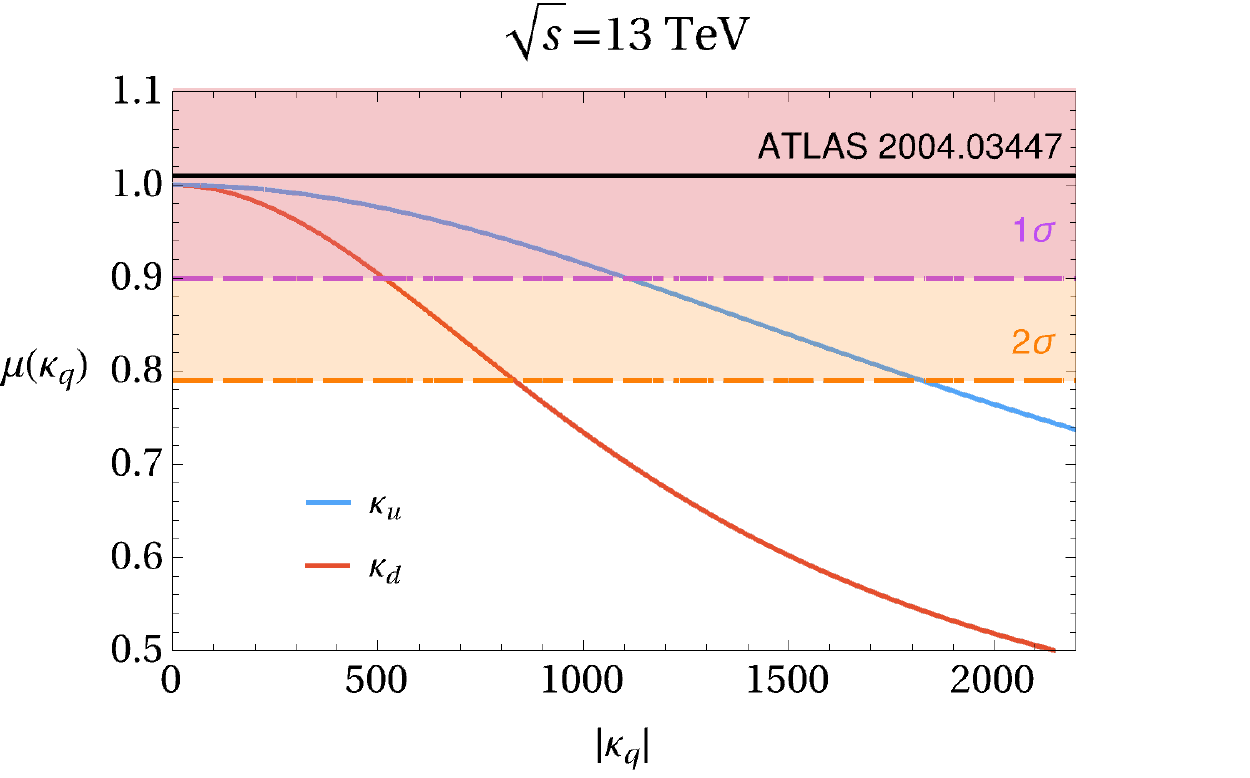} 
  \caption{Signal strengths for resonant Higgs production as a function of $\kappa_q$. The current measurement from ATLAS \cite{ATLAS:2020rej} is shown as a reference (black line), including the 1 and 2 $\sigma$ uncertainty bands.}
\label{fig:sig_str}
\end{figure}

We can now study the effect of the combination of all the ingredients discussed above for on-shell production, and we compare them to the SM case in fig.~\ref{fig:sig_str}, where we plot the signal strength
\begin{equation}
\mu(\kappa_q)= \frac{[\sigma (gg \to h)(\kappa_q) + \sigma (q \bar{q} \to h) (\kappa_q)] \cdot \text{BR}_{ZZ}(\kappa_q)}{\sigma (gg \to h)_\text{SM}  \cdot \text{BR}_{ZZ~ \text{SM}}}
\end{equation}
defined as the ratio of the NP on-shell production cross section, multiplied by the modified $h \to ZZ$ branching ratio, over the SM prediction (in which we omit the negligible $q\bar{q}$-initiated contribution). Under the assumption that the only NP modification is the separate enhancement of the down (up) Yukawa coupling, from a comparison with the latest ATLAS on-shell measurement \cite{ATLAS:2020rej} one can exclude values $\kappa_d \gtrsim 850 ~(\kappa_u \gtrsim 1850)$ at the $2\sigma$ confidence level, a constraint that is  around a factor of 3 weaker than the projected limits from a HL-LHC global fit \cite{deBlas:2019rxi}. 

Furthermore, as the enhancement via $\kappa_q$ has an impact both on the production cross section and on the decay width of the Higgs boson, the interplay of the two effects spoils the interpretation of eq.~\eqref{eq:onoff} as an indirect measurement of $\Gamma_h$. Finally, we note that on-shell measurements cannot be used as a model-independent probe of the first-generation quark Yukawa couplings, as the obtained limits rely on the strong assumption of light-quark Yukawa coupling modifications only, whereas e.g.~shifts in the Higgs coupling to $Z$ bosons would spoil the interpretation of the on-shell measurement. Therefore in the rest of the paper we move our attention to off-shell production.

\section{The off-shell Higgs \label{sec:offshell}}

In the case of off-shell Higgs production we cannot make use of the narrow width approximation as in eq.~\eqref{eq:nwa}, therefore in this section we reconsider the theoretical prediction for $pp \to ZZ$ and discuss how it is affected by an enhancement of the first-generation Yukawa couplings.

\begin{figure}
\centering
\begin{subfigure}{.5\textwidth}
  \centering
   \includegraphics[width=0.6 \linewidth]{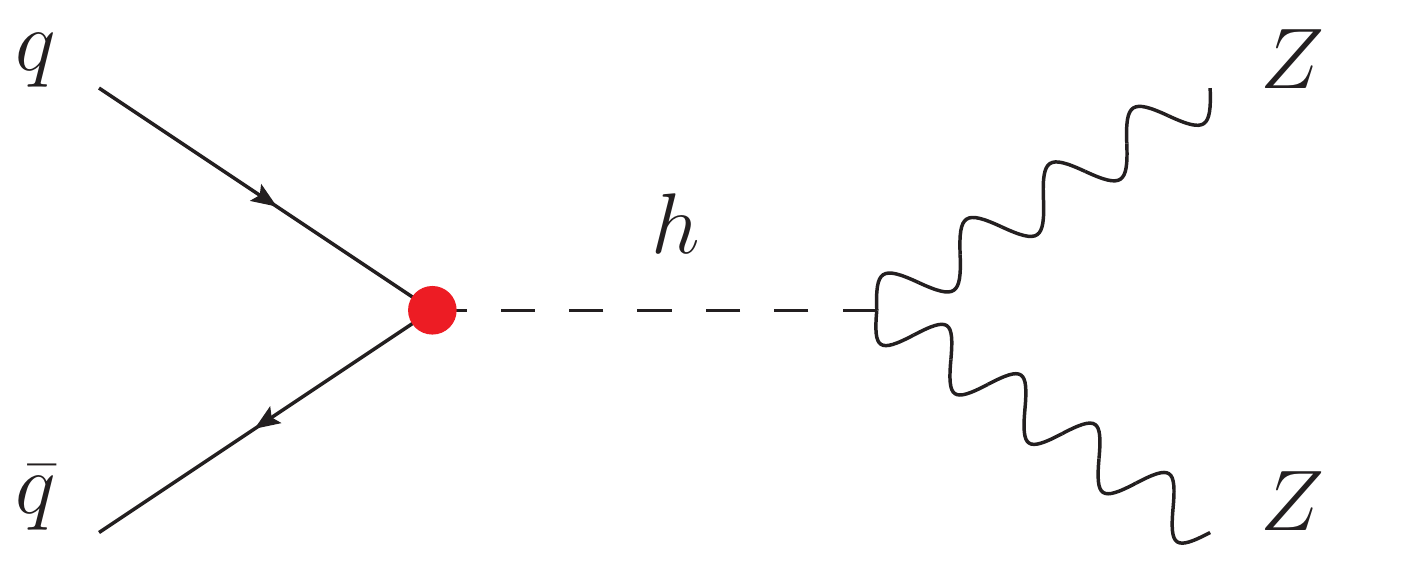} \\

  \caption{}
  \label{fig:qqh}
\end{subfigure}%
\begin{subfigure}{.5\textwidth}
  \centering
   \includegraphics[width=0.35 \linewidth]{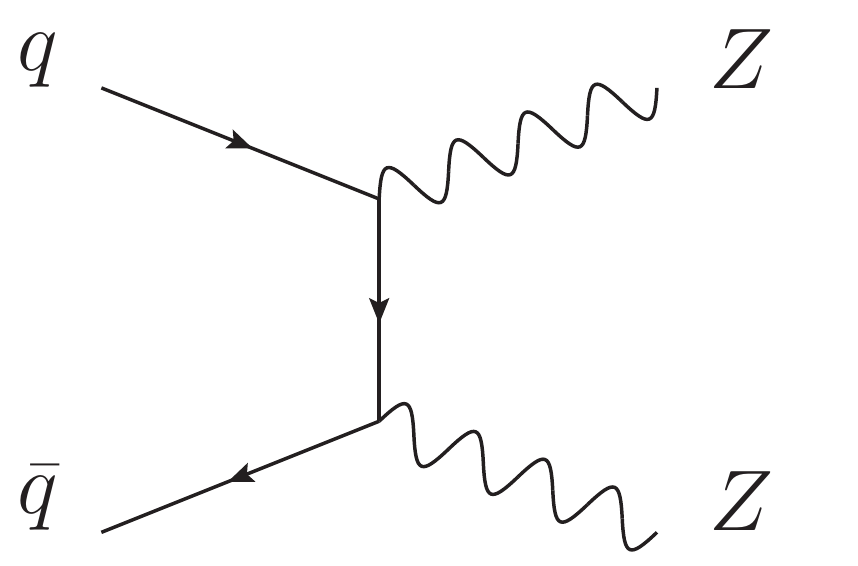} \\

  \caption{}
  \label{fig:qqt}
\end{subfigure}

\caption{Feynman diagrams contributing to $q\bar{q}\to ZZ$: (a) Higgs-mediated process, with the red dot indicating a modified light Yukawa coupling; (b) $\hat{t}$-channel diagram for the dominant production mode. }
\label{fig:diags}
\end{figure}

In refs.~\cite{Alasfar:2019pmn, Alasfar:2022vqw} it has been pointed out that light-quark Yukawa couplings can be constrained from Higgs pair production. This is mostly due to
an effective coupling of the light quarks to two Higgs bosons, which arises from the operators with Wilson coefficients $C_{u\phi}$ or $C_{d\phi}$, respectively. 
Thinking in terms of the Goldstone boson equivalence theorem, the same 
coupling gives a contribution to the longitudinal $Z$ boson pair production relevant for the off-shell region. 
The partonic differential cross section for the production of two neutral 
Goldstone bosons in the limit of large partonic centre-of-mass energy $\hat{s}$ is given by
\begin{equation}
\frac{d \hat \sigma_{q_i\bar{q}_j}}{d \hat t}= \frac{1}{16\pi}\frac{1}{3 \hat{s}} g_{G_0 G_0q_{i}q_{j}}^2\label{eq:GBequivalence}\,.
\end{equation}
The cross section for the quark-initiated off-shell Higgs production is 
hence expected to provide a potential probe of light-quark Yukawa 
couplings with a similar behaviour as for the Higgs pair production process, which
motivates a study of light-quark Yukawa couplings in the context of the off-shell Higgs measurements. Furthermore, we make use of the differences in the kinematics of the process with respect to the SM one.

\par
Similarly to what was observed in the previous section, the enhancement of the light Yukawa couplings implies that the Higgs-mediated quark-fusion channel (see fig.~\ref{fig:qqh}) must be taken into account when considering off-shell Higgs production. The LO partonic differential cross section for $q\bar{q}\to h^*\to ZZ$ is given by 
\begin{align}
\frac{d \hat \sigma_{q_i\bar{q}_j}}{d \hat t} &= g_{h q_i\bar{q}_j}^2\frac{1}{16 \pi}\, \frac{1}{12  \hat{s}} \frac{1}{v^2} \frac{1}{(\hat{s}-m_h^2)^2}\bigg[ 12 m_Z^4-4 m_Z^2 \hat{s} +\hat{s}^2 \bigg],
\label{eq:dsigqqdt}
\end{align}
which in the limit of large $\hat{s}$ corresponds to eq.~\eqref{eq:GBequivalence}. 
From this the hadronic cross section can be obtained by
\begin{equation}
  \sigma_{\mathrm{hadronic}} =  \int_{\tau_0}^1 d\tau \int_{\hat{t}_-}^{\hat{t}_+} 
  d\hat{t} \sum_{i,j} \frac{d\mathcal{L}^{q_i\bar{q}_j}}{d\tau}\frac{ d\hat \sigma_{q_i\bar{q}_j}}{d \hat t}\,,
\end{equation}
with $ \tau_0= 4\, m_Z^2/s$, $\hat{s}=\tau s$ and
\begin{equation}
\hat{t}_{\pm}=m_Z^2-\frac{\hat{s}(1\mp \beta)}{2} \quad\quad 
\text{and}\quad \quad \beta=\sqrt{1-\frac{4 m_Z^2}{\hat{s}}}\,.
\end{equation}
The parton luminosity is given by
\begin{equation}
  \frac{d{\cal L}^{q_i \bar q_j}}{d\tau} = \int_\tau ^1 \frac{dx}{x} \,\left[  f_{q_i}(\tau/x,\mu_F^2) f_{\bar{q}_j}(x,\mu_F^2) + \,f_{\bar{q}_j}(\tau/x,\mu_F^2) f_{q_i}(x,\mu_F^2)\right]\,.
\end{equation}
We neglected all the kinematical masses of the light quarks, while the coupling of the Higgs boson to the light quarks (for flavour diagonal couplings) is given by eq.~\eqref{eq:ghqq} or eq.~\eqref{eq:ghqqkappa}. 

\begin{figure}
\centering
\begin{subfigure}{.5\textwidth}
  \centering
   \includegraphics[width=0.6 \linewidth]{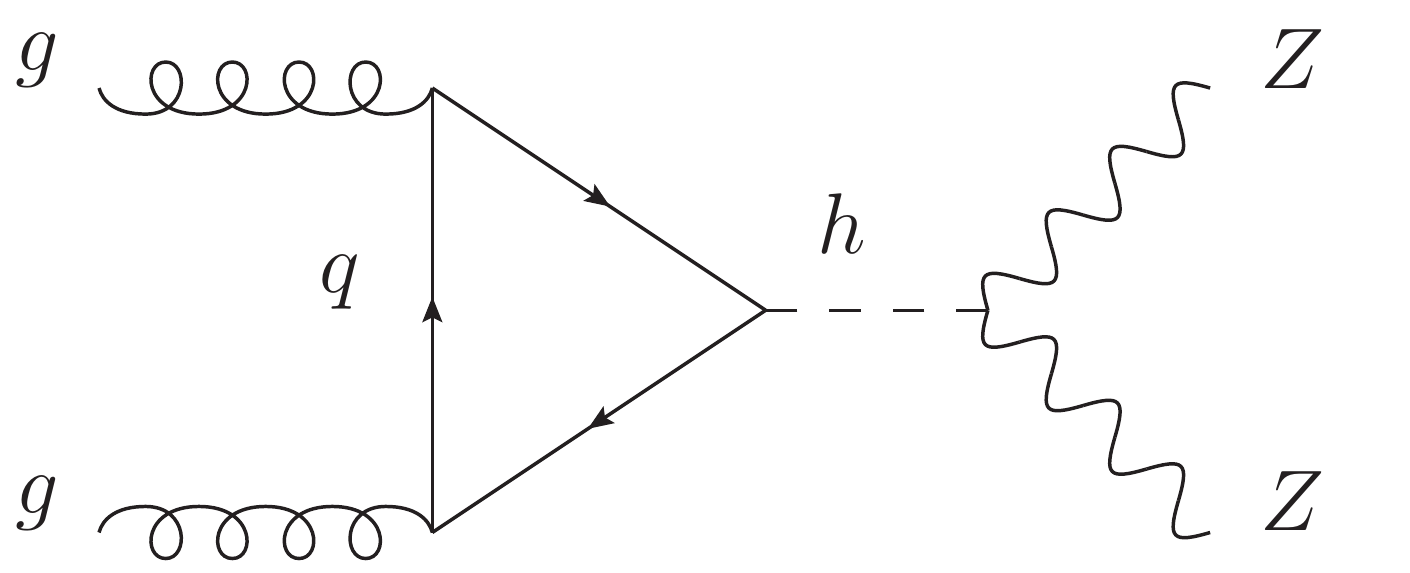} \\

  \caption{}
  \label{fig:tri}
\end{subfigure}%
\begin{subfigure}{.5\textwidth}
  \centering
   \includegraphics[width=0.6 \linewidth]{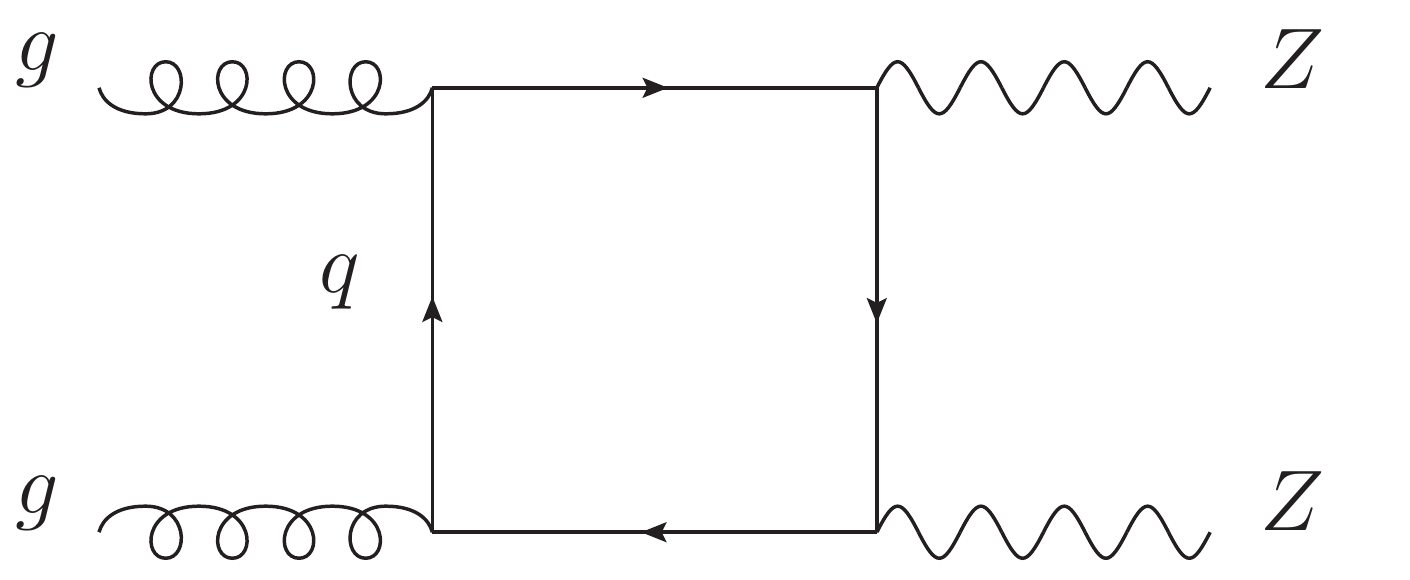} \\

  \caption{}
  \label{fig:box}
\end{subfigure}

\caption{The two topologies for the Feynman diagrams occurring in the LO contribution to $gg\to ZZ$: (a) triangles associated to the Higgs-mediated process; (b) boxes associated to continuum production.}
\label{fig:diags}
\end{figure}

Within the SM an off-shell Higgs boson is produced via gluon fusion, where a loop of third-generation quarks\footnote{We recall that we are neglecting the effect of a modified $\kappa_q$ on the gluon-fusion channel.} couples to an off-shell Higgs boson, as in fig.~\ref{fig:tri}. This process has large interference with box diagrams from \emph{continuum} production, $gg\to ZZ$, see fig.~\ref{fig:box}. We used the results of ref.~\cite{Vitti:2022kot} for the gluon-fusion cross section at LO, including the Higgs-mediated, the continuum and the interference contributions.

Finally, the main background to off-shell Higgs production is $q\bar{q}\to ZZ$ production, occurring via diagrams as shown in fig.~\ref{fig:qqt}. We computed the LO cross section and found agreement with the result of ref.~\cite{Mele:1990bq}. We note that there is no signal -- background interference between the amplitudes for $q\bar{q}\to h^*\to ZZ$ and $q\bar{q}\to ZZ$. 
In particular, using the spinor-helicity formalism, one can observe that the $q\bar{q}\to h^*\to ZZ$ amplitude receives a non-zero contribution only when the (massless) fermions in the initial state have the same helicities, whereas for $q\bar{q}\to ZZ$ the initial fermions must have opposite helicities. For this reason the signal and the background processes are related to different helicity amplitudes, and cannot interfere. We remark that this helicity selection rule holds at higher perturbative orders, and that it is insensitive to the polarisation of the $Z$ bosons in the final state, so that the interference cannot be \textit{resurrected} e.g.~by looking at the decay leptons of the $Z$ bosons \cite{Panico:2017frx, Azatov:2017kzw}.
However, the above reasoning does not automatically imply that, in the perturbative expansion of the SMEFT, we are considering purely BSM, $\mathcal{O}(1/\Lambda^4)$, effects. Indeed, the BSM/SM interference is already contained in the square of $g_{h q_i\bar{q}_j}$ in eq.~\eqref{eq:dsigqqdt}, which in fact includes both $\mathcal{O}(1/\Lambda^2)$ and $\mathcal{O}(1/\Lambda^4)$ contributions (see eq.~\eqref{eq:ghqq}). Still, the $\mathcal{O}(1/\Lambda^2)$ interference is very small because of the smallness of the light quark masses in the SM, and in the rest of the paper we will neglect it. We note that the fact that our bound is dominated by $\mathcal{O}(1/\Lambda^4)$ effects does not invalidate the analysis given that a potential contribution of a dimension-8 operator would again be suppressed by the small quark masses.

\section{Phenomenological analysis \label{sec:analysis}}
In the following we analyse the sensitivity of the HL-LHC on the light-quark Yukawa coupling modifications from measurements of off-shell Higgs production. The signal processes are $d\bar{d}\to h^* \to ZZ\to 4\ell$  and $u\bar{u}\to h^* \to ZZ\to 4\ell$ with enhanced Yukawa couplings respectively, while the background processes are given by gluon fusion and the quark-induced $ZZ$ production. 
We consider on-shell $Z$ bosons only, which is a very good description 
if the invariant mass of the $Z$ boson pair is sufficiently above the kinematic threshold \cite{Campbell:2014gua}. We impose a cut of $m_{ZZ}>250\text{ GeV}$ to assure the accuracy of this simplification. We reproduce the effect of basic selection cuts on the transverse momentum of the leptons, $p_{T\ell}> 10\text{ GeV}$, and on the pseudorapidity of the leptons, $|\eta_{\ell}|<2.5$, via efficiency factors, obtained using $\texttt{MadGraph\_aMC@NLO}$ \cite{Alwall:2014hca}.
\par
In the analysis we use the LO matrix elements introduced in sec.~\ref{sec:offshell} but we improve them with $K$-factors. Currently, the background of four-lepton production via $q\bar{q}$ annihilation is known at next-to-next-to-leading order (NNLO) \cite{Cascioli:2014yka, Grazzini:2015hta, Heinrich:2017bvg}, for which we use a $K=\sigma_{NNLO}/\sigma_{LO}=1.6$.
The NLO QCD corrections on the signal can be inferred from $b\bar{b}h$ production \cite{Dicus:1998hs, Balazs:1998sb, Harlander:2003ai}. We find $K^{u}=1.47$ and $K^d=1.63$ for the up- and down-initiated channels for a scale choice of $\mu_R=\mu_F=M_{ZZ}/2$.
The gluon-fusion contribution is fully known only at LO \cite{PhysRevD361570, GLOVER1989561}. The NLO QCD corrections to the Higgs-mediated diagrams \cite{Aglietti:2006tp, Anastasiou:2006hc, Harlander:2005rq} and to the continuum production via loops of massless quarks \cite{vonManteuffel:2015msa, Caola:2015psa} are known in analytic form; the two-loop diagrams mediated by top-quark loops have been computed numerically in refs.~\cite{Agarwal:2020dye, Bronnum-Hansen:2021olh} but they are still not included in a full NLO prediction.
We use a $K=1.83$ for the gluon fusion background following the estimate in refs.~\cite{Haisch:2021hvy, Buonocore:2021fnj}. We use the {\tt NNPDF40\_lo\_as\_01180} parton distribution functions \cite{NNPDF:2021njg} and $\sqrt{s}=14\text{ TeV}$.
\par
\begin{figure}
\begin{subfigure}{.5\textwidth}
  \centering
  \hspace*{-1cm}
   \includegraphics[width=1.1 \textwidth]{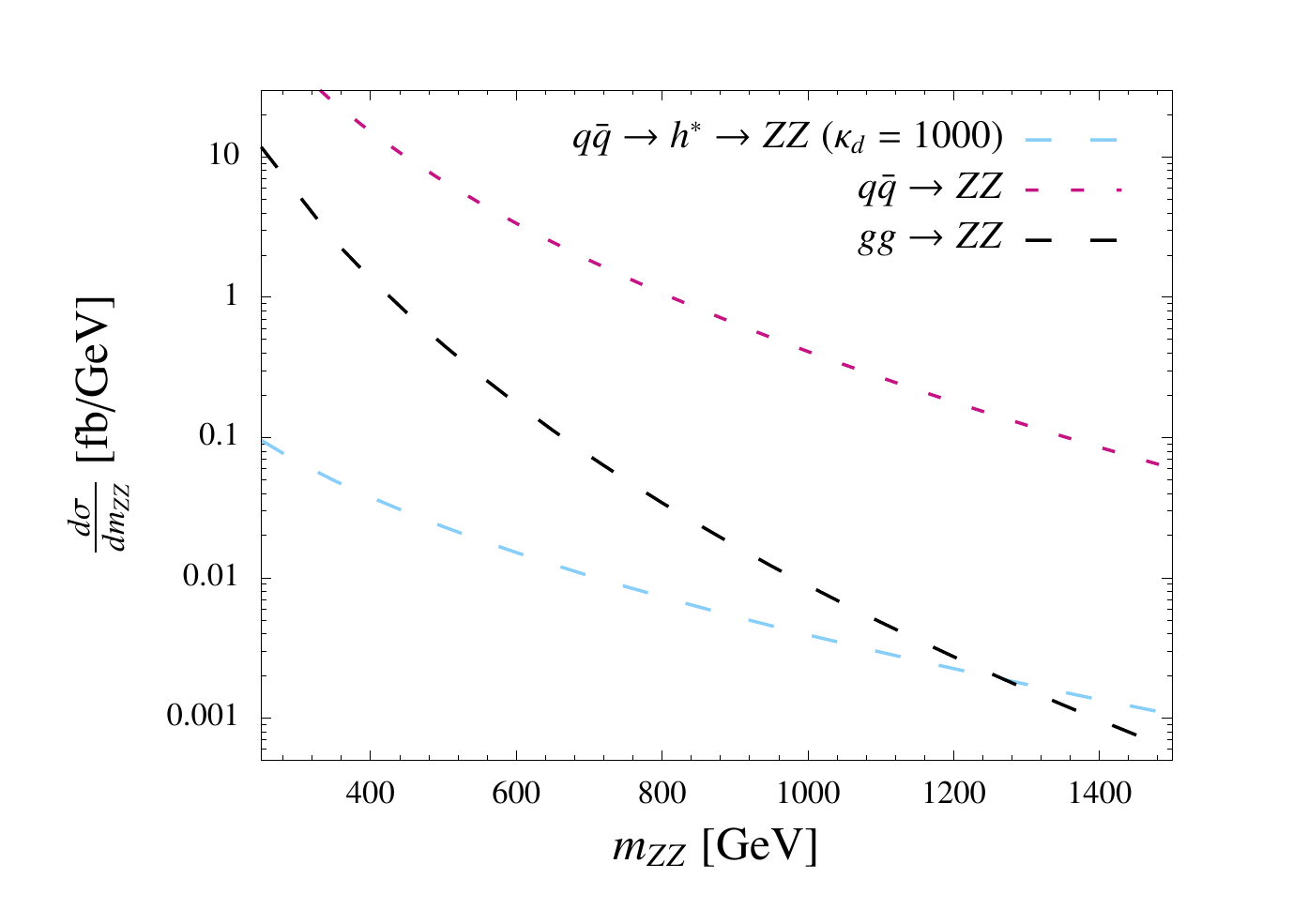}  \\

  \caption{}
  \label{fig:mzzdist}
\end{subfigure}%
\begin{subfigure}{.5\textwidth}
  \centering
  \hspace*{-0.5cm}
   \includegraphics[width=1.1 \textwidth]{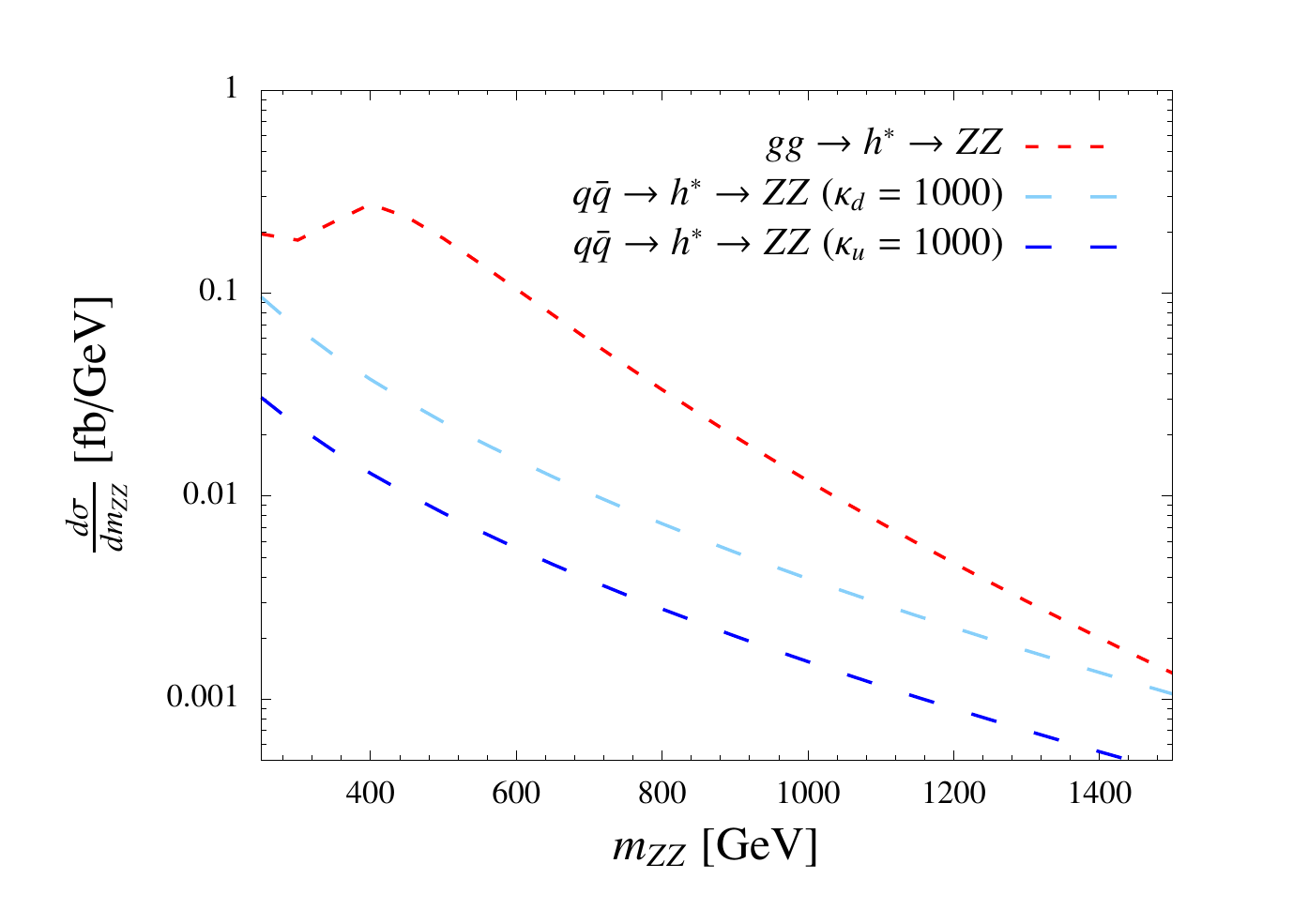} \\

  \caption{}
  \label{fig:mzzdist_tri}
\end{subfigure}

\caption{Distribution in the invariant mass $m_{ZZ}$ for the various $ZZ$ production channels. (a) The $q\bar{q}\to ZZ$ SM background (violet), the $gg\to ZZ$ background (black) and the signal $d\bar{d}\to h^* \to ZZ$ (light blue) for a value $\kappa_d=1000$, which corresponds to $\tilde{C}_{d\phi}/(1 ~\text{TeV}^2)= 0.45$. (b) Comparison of the $gg\to h^* \to ZZ$ (triangle) contribution to the background with the $d\bar{d}\to h^* \to ZZ$ signal (light blue) and the $u\bar{u}\to h^* \to ZZ$ (dark blue) for $\kappa_u=1000$ which corresponds to $\tilde{C}_{u\phi}/(1 ~\text{TeV}^2)=0.21$. }
\label{fig:mZZ}
\end{figure}

In fig.~\ref{fig:mZZ} we show the invariant mass distribution for signal and background processes. For the signal we chose $\kappa_u=\kappa_d=1000 $, corresponding to $\tilde{C}_{u\phi}= 0.21$ and $\tilde{C}_{d\phi}=0.45$ for a NP scale fixed at $\Lambda=1~\text{TeV}$. While the $q\bar{q}$ background is about two orders of magnitude larger than the signal, we can see that the latter contribution wins relative importance over the gluon-fusion background in the high invariant mass bins. This is a consequence both of the $\kappa_q$ enhancement and of the different fall of the quark PDFs compared to the gluon one. Furthermore, the destructive interference between the Higgs-mediated and continuum amplitudes for ggF at high invariant masses plays a role in enhancing the $q\bar{q}$ signal over the ggF background (compare figs.~\ref{fig:mzzdist} and \ref{fig:mzzdist_tri}).

\par
For the measurement of the Higgs width kinematic discriminants based on the matrix element method have proven to be very powerful \cite{Campbell:2013una, ATLAS-CONF-2022-068, CMS:2022ley}. While the matrix element method is usually complicated by the fact that it requires the knowledge of a transfer function that describes the transfer of an event with parton-level momentum to an event with smeared detector-level momentum, for the off-shell Higgs analysis these terms cancel for the clean four-lepton final state when employing discriminants based on ratios of signal and background matrix elements. We define the weighted probability for an event initiated by the partons $i,j$, and with a fixed set $v$ of kinematic variables, as
\begin{equation}
P_{ij}(v) =\frac{1}{\sigma_{ij\to 4 \ell}} \int d x_1 d x_2 \delta(x_1 x_2 E_{CMS}^2 - m_{4\ell}^2) f_i(x_1) f_j(x_2)\hat{\sigma}_{ij}(x_1, x_2, v)\,,
\label{eq:probab}
\end{equation}
where $f_i$ and $f_j$ are the parton distribution functions, $E_{CMS}$ is the collider energy and $\sigma_{ij\to 4 \ell}$ is the hadronic cross section for the process initiated by $i$ and $j$.
We can then define the kinematic discriminants
\begin{equation}
D_s^d=\log_{10}\left( \frac{P_{d\bar{d}}^{sig}}{P_{q\bar{q}}^{back}+P_{gg}^{back}}\right) \quad \quad \text{and} \quad \quad D_s^u=\log_{10}\left( \frac{P_{u\bar{u}}^{sig}}{P_{q\bar{q}}^{back}+P_{gg}^{back}}\right)\,.
\label{eq:defds}
\end{equation}
We note that the dependence on the enhanced light-quark Yukawa couplings drops out in the respective definitions of $P_{q \bar{q}}^{sig}$. We show our results for $D_s^d$ in this section while the results based on $D_s^u$ are shown in appendix~\ref{app:Dsu}. 
 \begin{figure}
\centering  \includegraphics[width= 0.8 \textwidth]{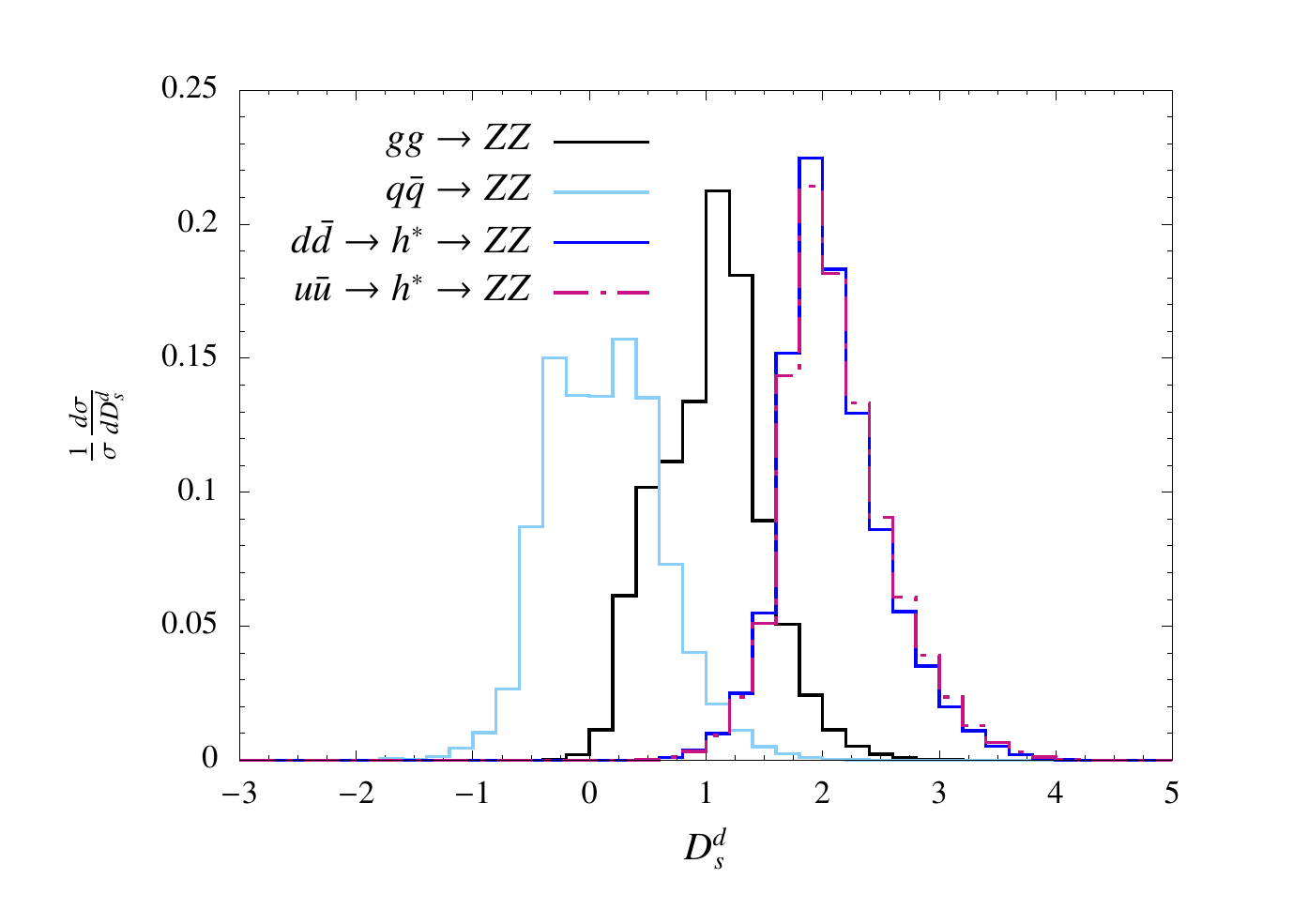} 
     \caption{Normalised differential distributions with respect to $D_s^{d}$ for signal (blue and pink dashed) and background (light blue and black) processes.  \label{fig:distDS} }
 \end{figure}
  \begin{figure}
\centering  \includegraphics[width= 0.8 \textwidth]{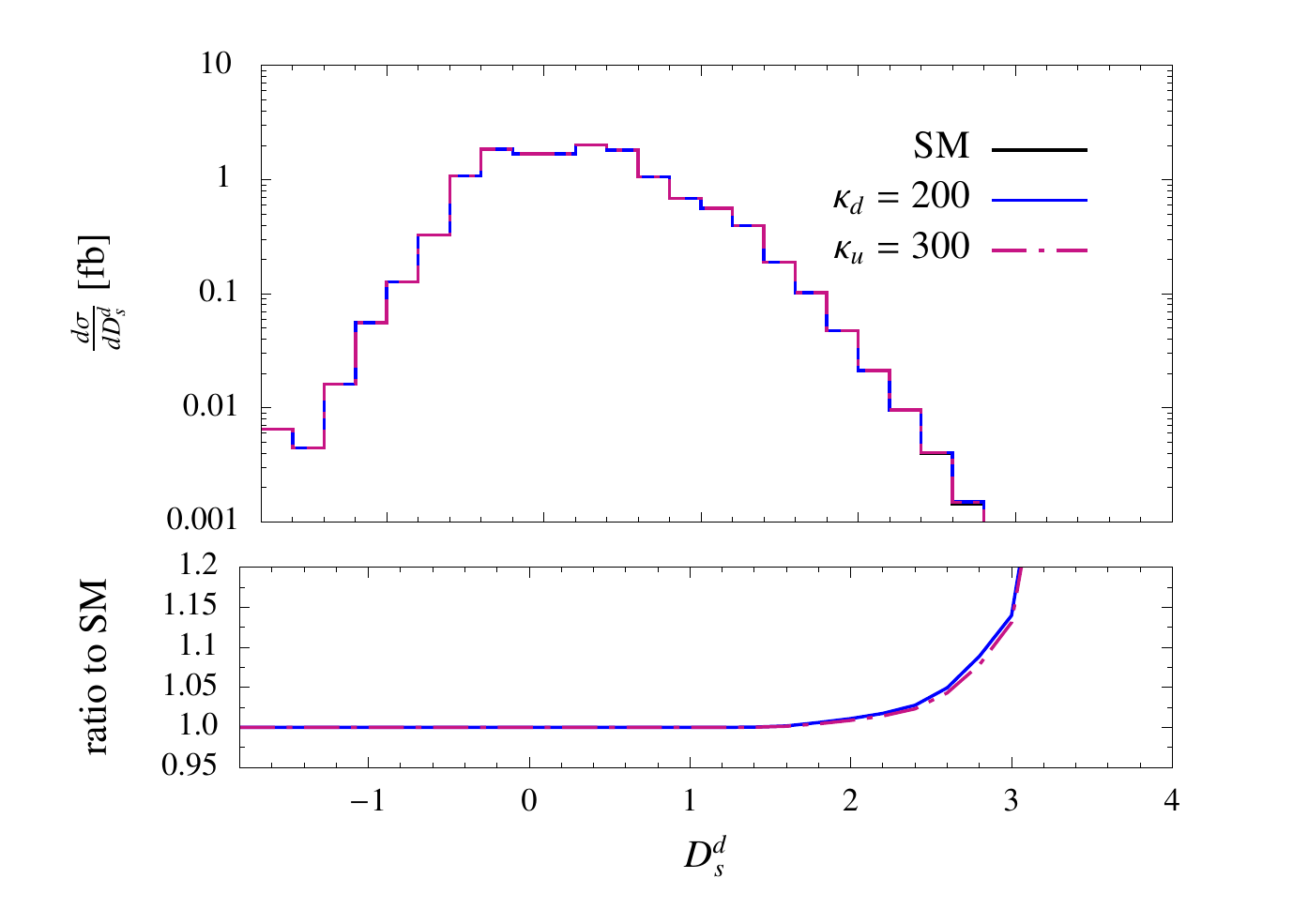} 
     \caption{Distribution of the cross section of SM (black), SM+signal with $\kappa_d=200$ (blue) and SM+signal with $\kappa_u=300$ (pink dashed) as function of $D_s^d$. The lower panel shows the ratio to the SM.  \label{fig:ratio}}
 \end{figure}
In fig.~\ref{fig:distDS} we show the normalised distribution of the signal processes $d\bar{d}\to h^*\to ZZ$ as a blue line and $u\bar{u}\to h^*\to ZZ$ as a pink dashed line, while the $gg\to ZZ$ background is shown as a black line and the $q\bar{q}\to ZZ$ background is shown in light blue. The figure clearly demonstrates the discriminating power of the $D_s^d$ variable, stemming mainly from the bins $D_s^d > 2$, as confirmed by fig.~\ref{fig:ratio}, where we compare the expected $D_s^d$ distributions for the $pp \to ZZ \to 4\ell$ cross section in the SM and in the NP scenarios with $\kappa_d=200$ and $\kappa_u=300$. The $d\bar{d}$- and $u\bar{u}$-induced signals show instead very similar  $D_s^d$ distributions, hence the process will likely not allow to distinguish between an enhanced up or down Yukawa hypothesis. We note though that this could be done in a global fit, using for instance also limits from $h\gamma$ where the up- and down-Yukawa contributions are distinguished due to the different quark charges \cite{Aguilar-Saavedra:2020rgo}. 
We have also investigated the use of a kinematic discriminant as defined in the analysis in ref.~\cite{ATLAS:2019qet}\footnote{While we find the same quantitative and qualitative behaviour of the thus defined $D_s$ we find a shift on the $x$ axis that we attribute to a different normalisation.}, and found that while it leads to slightly worse limits on the light-quark Yukawa couplings compared to our definition of $D_s$, it still shows very good discriminating power. The experimental analyses might hence be sensitive to similar order of magnitude modifications of light-quark Yukawa couplings compared to what we find in our study, without implementing a dedicated analysis. 

\begin{figure}
\centering
\hspace*{0cm}
 \includegraphics[width=0.5 \textwidth]{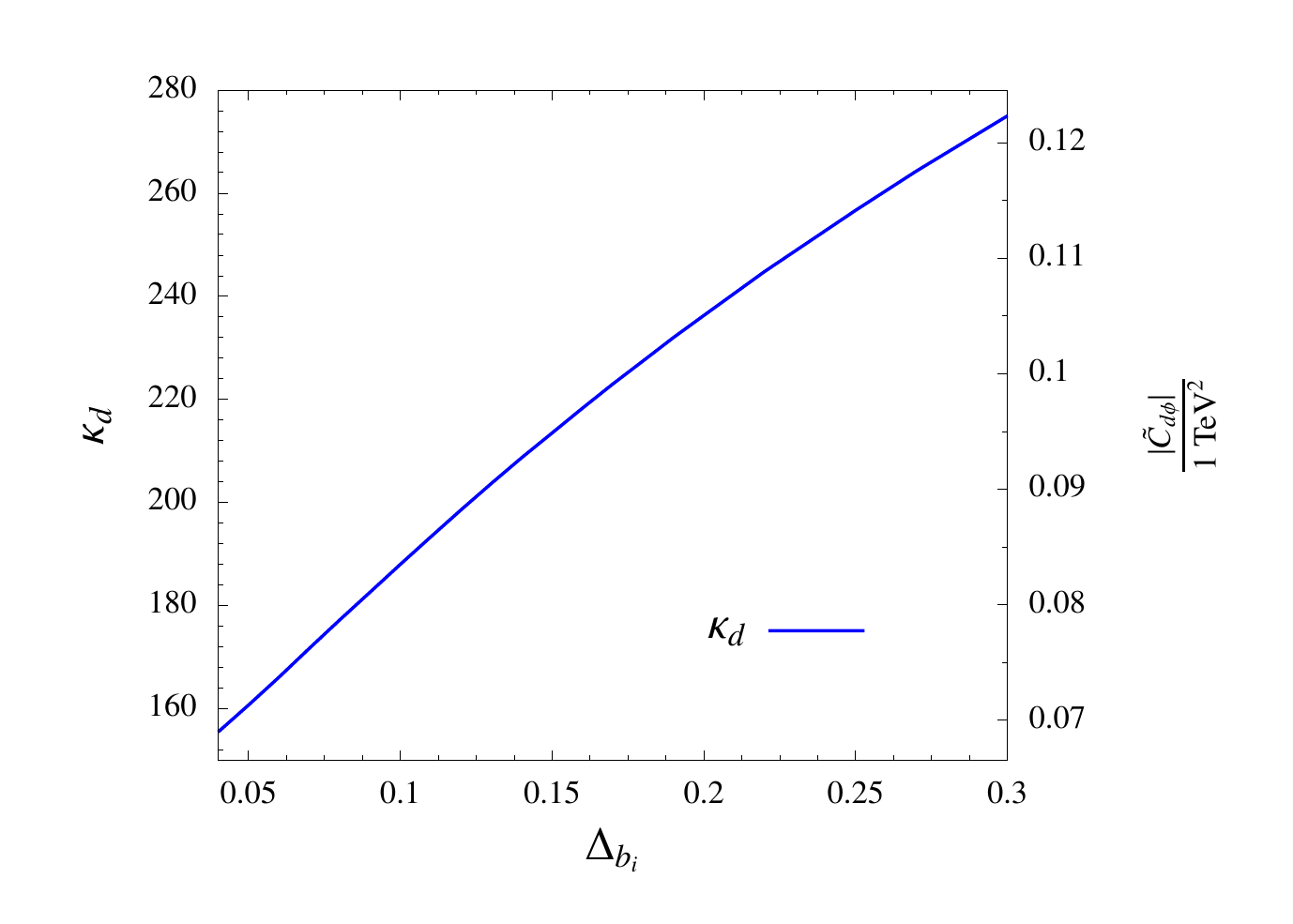} 
\hspace*{-0.5cm}
   \includegraphics[width=0.5 \textwidth]{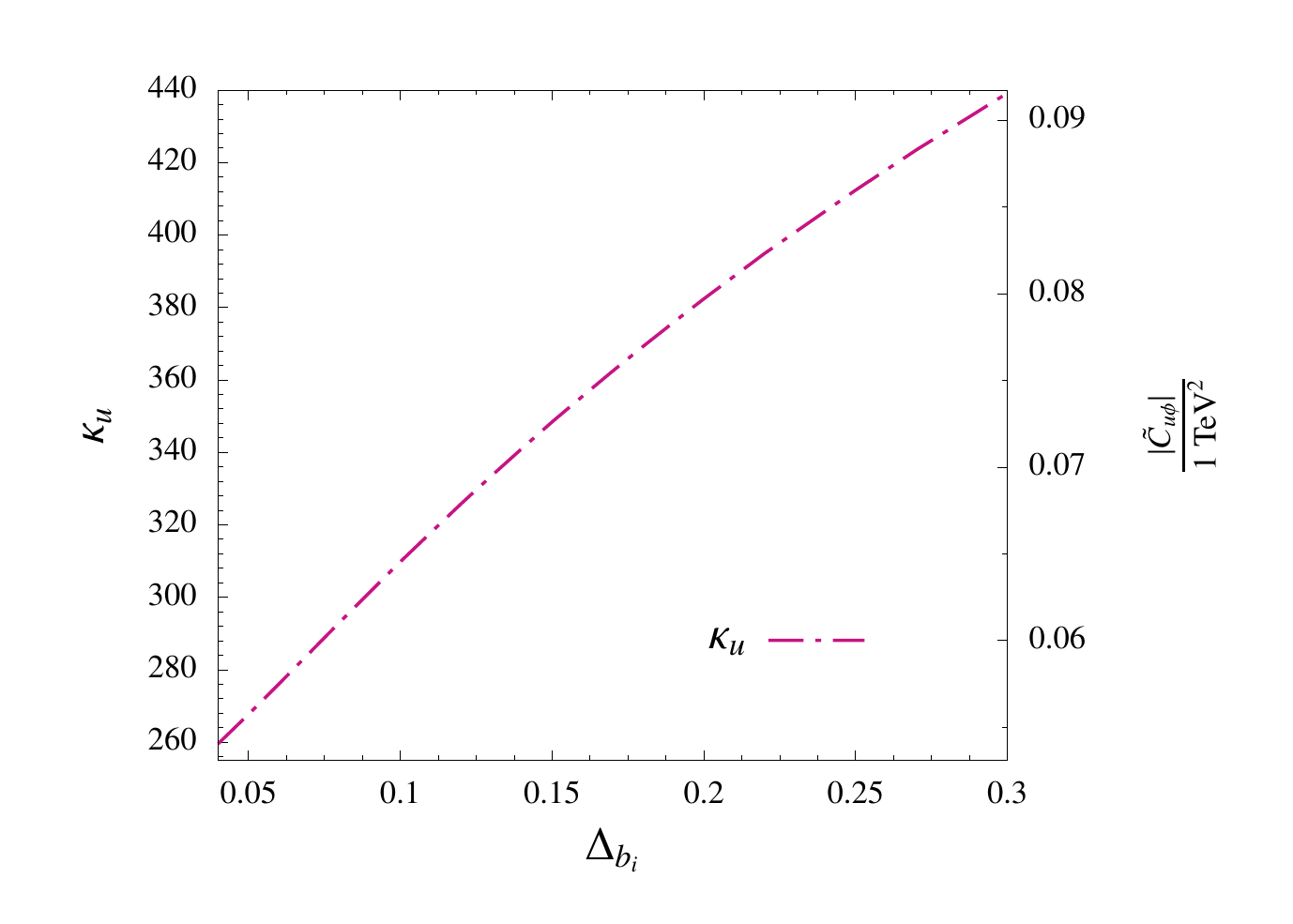} \hspace*{-0.5cm}
 \caption{Dependence of the sensitivity bounds on $\kappa_d$ ($\tilde{C}_{d\phi}$) in the left panel and $\kappa_u$ ($\tilde{C}_{u\phi}$) in the right panel on the assumption made on the size of $\Delta_{b_i}$.  \label{fig:sigmadep}}
\end{figure}

In order to set limits on the light-quark Yukawa couplings we perform a shape analysis on the $D_s^d$ distributions. While we could in principle also include the $m_{ZZ}$ distribution in the analysis we found no difference doing so. The significance in the $i$-th bin is computed as a Poisson ratio of likelihoods that incorporates uncertainties on the background using the Asimov approximation \cite{Cowan:2010js}
\begin{equation}
Z_i=\sqrt{2\left[(s_i+b_i )\ln\frac{(s_i+b_i)(b_i+\sigma_{b_i}^2)}{b_i^2+(s_i+b_i)\sigma_{b_i}^2} -\frac{b_i^2}{\sigma_{b_i}^2} \ln\left( 1+\frac{s_i \sigma_{b_i}^2}{b_i(b_i+\sigma_{b_i}^2)}\right)\right]}\,,
\end{equation}
where $s_i$ and $b_i$ are respectively the number of signal and background events in the $i$-th bin. The standard deviation $\sigma_{b_i}=\Delta_{b_i} b_i$ characterises the (experimental and theoretical) uncertainties of the associated background in the bin. We assume a flat uncertainty and show in fig.~\ref{fig:sigmadep} the dependence of the sensitivity limit on $\kappa_d$ and $\kappa_u$ ($\tilde{C}_{d\phi}$ and $\tilde{C}_{u\phi}$) in dependence of $\Delta_{b_i}$. Concerning our choice for the range of the plot, the lower limit is based on the total experimental systematic uncertainties expected for the $gg\to h^* \to ZZ$ signal strength, amounting to 5.0\% and 3.9\% in the baseline scenarios S1 and S2, respectively, as given by the ATLAS Collaboration \cite{ATL-PHYS-PUB-2018-054}. The corresponding uncertainties of  the CMS Collaboration are 7.3\% and 4.1\% \cite{CMS-PAS-FTR-18-011}. Instead the upper limit  $\Delta_{b_i}=0.3$ corresponds to the approach advocated in ref.~\cite{Haisch:2021hvy}, motivated by the observation that the scale uncertainties do not capture the difference between LO and $K$-factor improved prediction for $pp \to ZZ$. For this reason ref.~\cite{Haisch:2021hvy} proposed to take instead half of the difference between LO and $K$-factor improved prediction. Compared to the combination of scale and PDF+$\alpha_s$ uncertainty this approach leads to a much larger uncertainty and can hence be considered as very conservative.

We find that, assuming the optimistic scenario of $\Delta_{b_i}=0.04$  at the HL-LHC, it is possible to obtain the following constraints
\begin{align*}
|\tilde{C}_{d\phi}|/(1 \text{ TeV})^2< 0.069/\text{TeV}^2 \quad (\kappa_d< 156 ), \\
 |\tilde{C}_{u\phi}|/(1 \text{ TeV})^2< 0.054/\text{TeV}^2 \quad (\kappa_u<  260).
\end{align*} 
 \begin{figure}[h!]
\centering  \includegraphics[width= 0.7 \textwidth]{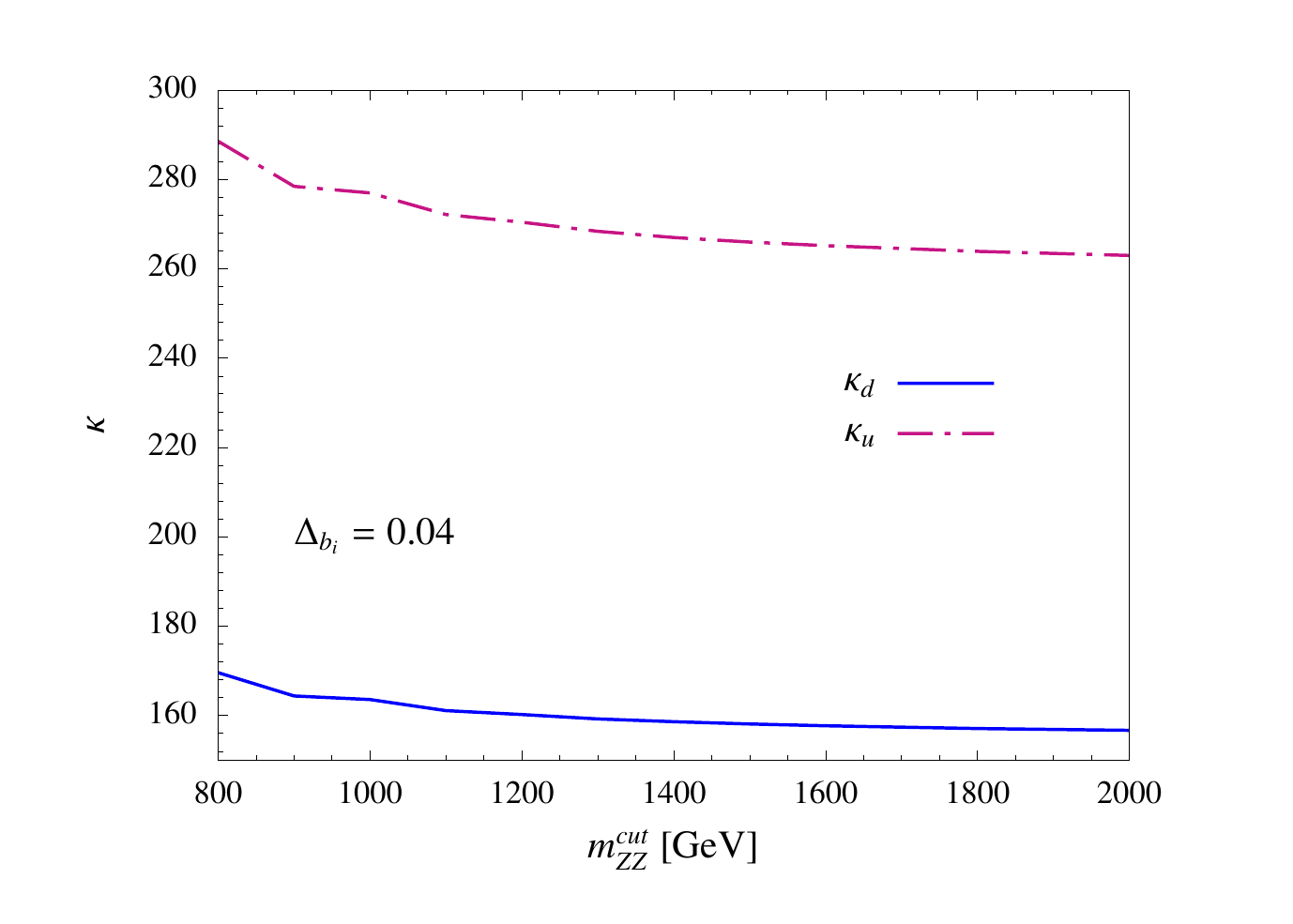} 
     \caption{The projected sensitivity on $\kappa_d$ (blue line) and $\kappa_u$ (pink dashed line) as a function of an upper cut $m_{ZZ}< m_{ZZ}^{cut}$ using $\Delta_{b_i}=0.04$. \label{fig:slicing} }
 \end{figure}
In fig.~\ref{fig:slicing} we show the dependence of the sensitivity limit on $\kappa_d$ and $\kappa_u$ on an upper cut on the invariant $ZZ$ mass, hence imposing $m_{ZZ}< m_{ZZ}^{cut}$, in order to check the validity of our EFT approach. The procedure of providing limits in terms of an upper cut on the energy probed, a.k.a.~\textit{clipping}, was recommended in ref.~\cite{Brivio:2022pyi}. We emphasise, given that the SM first-generation Yukawa couplings can be neglected, that the sensitivity in our analysis on the coefficients $\tilde{C}_{u\phi}$ or $\tilde{C}_{d\phi}$ (or $\kappa_d$ and $\kappa_u$) stem purely from the dimension-six squared terms. Furthermore, in regard of fig.~\ref{fig:slicing} we conclude that the sensitivity of our analysis does not stem only from bins of high invariant mass, hence the EFT approach seems appropriate.
\par
 \begin{figure}
\includegraphics[width= 0.9 \textwidth]{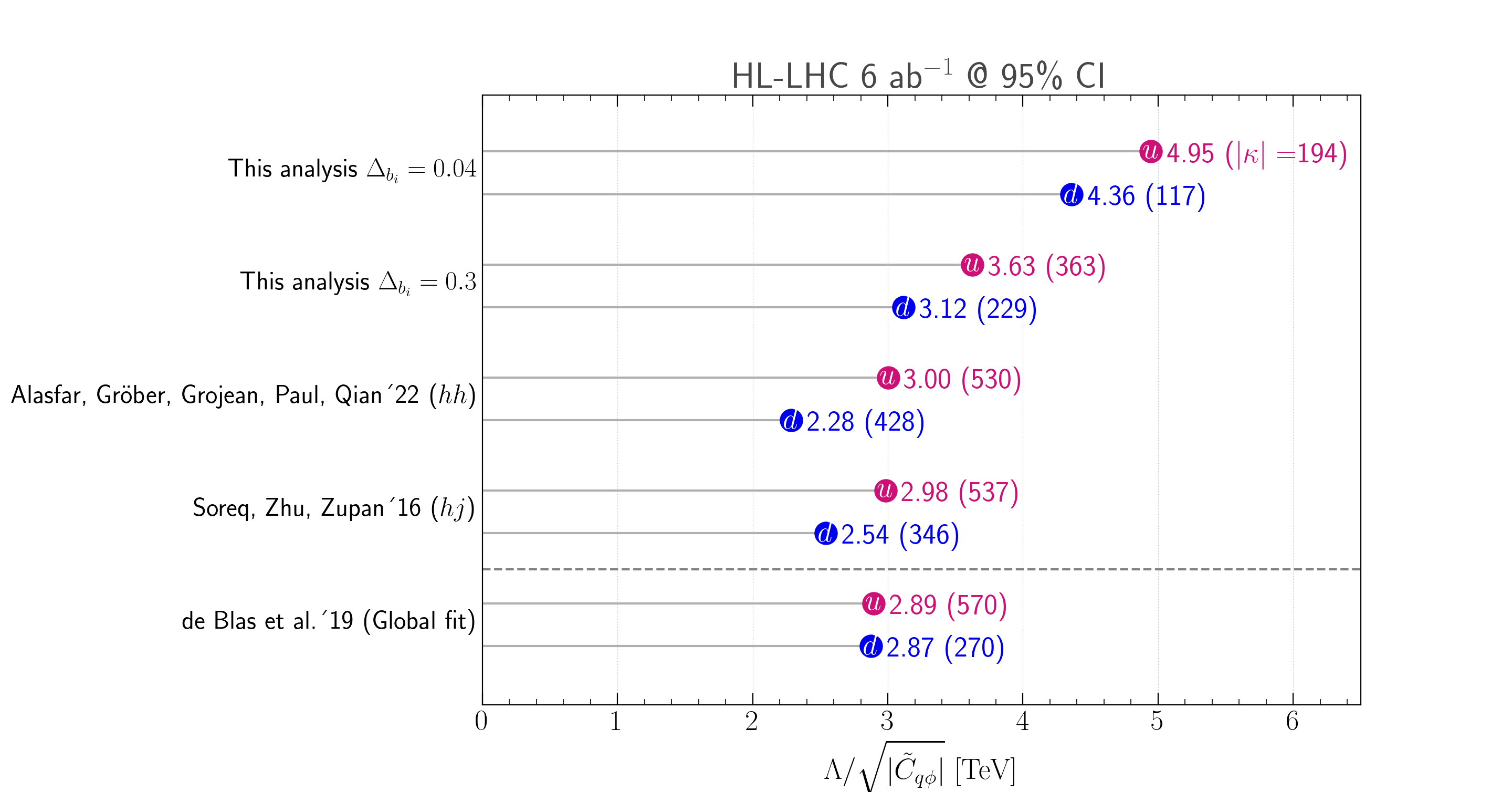} 
     \caption{Comparison of projected constraints on $\tilde{C}_{d\phi}$ (blue) and $\tilde{C}_{u\phi}$ (pink) at 95\% confidence level for the HL-LHC with $6~\text{ab}^{-1}$ luminosity, as estimated by this paper and previous analyses \cite{Alasfar:2022vqw, Soreq:2016rae, deBlas:2019rxi}. The constraints are interpreted in terms of the NP scale $\Lambda$ that can be probed via the measurement of the Wilson coefficients. The corresponding bounds on $\kappa_q$ are included in parentheses.}
\label{fig:summary}
 \end{figure}
Finally, we would like to emphasise that the off-shell Higgs measurement so far seems to provide the most sensitive probe of both the up- and the down-quark Yukawa couplings. In fig.~\ref{fig:summary} we compare our results with the ones obtained from alternative probes of the light-quark Yukawa couplings, \textit{cf.} the summary plot of ref.~\cite{Alasfar:2022vqw}. Even the constraints obtained in the conservative scenario assuming $\Delta_{b_i}=0.3$ are still competitive with those found in ref.~\cite{Soreq:2016rae}.

\section{Conclusion \label{sec:con}}
Because of their smallness in the SM, the light-quark Yukawa couplings are notoriously difficult to probe. In this paper we have studied the potential of off-shell Higgs production for a measurement of the first-generation quark Yukawa couplings. We decouple via an EFT language the quark masses from the couplings. This approach is particularly useful when dealing with scenarios involving large Yukawa couplings and massless quarks. The small effective couplings resulting from the decoupling process validate the use of perturbation theory and the application of a dimension-6 EFT analysis.
\par
Large enhancements of the up and down Yukawas lead to an increase
of the Higgs total width, which reduces the Higgs branching ratios in its standard decay
channels. On the other hand, large enhancements open up a new Higgs production channel, where the 
Higgs boson is directly produced via annihilation of the partons. 
The interplay of these effects prevents that the off-shell measurement can be interpreted in 
terms of a measurement of the Higgs width in the context of such scenarios. 
Instead, the off-shell signal strength is modified, and using appropriate kinematic discriminants allows
to give sensitivity limits on the modifications of the light-quark Yukawa couplings. 
Indeed, the signal and background amplitudes exhibit distinct characteristics (e.g.~no interference related to the helicity selection rule), providing strong discriminating power. The off-shell region provides enhanced signal contributions due to the behaviour of the total cross section. This enhancement is particularly notable for processes involving the longitudinal $Z$ boson. In the case of ggF, a similar enhancement in the off-shell region occurs. However, at high energy scales, the decay of the quark luminosity is slower compared to the gluon luminosity. This disparity in the energy behaviour results in an increased relative contribution from the quark channel in the off-shell region.

In our analysis we have found that off-shell Higgs production is a very promising channel to constrain the light-quark Yukawa couplings: using an optimistic
scenario with systematic uncertainties only of 4\% we find that $\kappa_d> 156$ and $\kappa_u > 260$ can be excluded at the HL-LHC, where $\kappa_q$ is the modification factor of the SM Yukawa coupling evaluated for $m_u=2.2\text{ MeV}$ and $m_d=4.7\text{ MeV}$. 
These projected sensitivities are better than the ones obtained studying other processes, such as Higgs pair production \cite{Alasfar:2019pmn, Alasfar:2022vqw}, Higgs+jet \cite{Soreq:2016rae}, Higgs+photon \cite{Aguilar-Saavedra:2020rgo}, $VVV$ \cite{Falkowski:2020znk} or the charge asymmetry in $W^{\pm}h$ \cite{Yu:2017vul}.
We have also considered how the projected sensitivity depends on the assumption made on the uncertainties, and we have seen that the limits from off-shell production are competitive with other approaches even in a conservative scenario.

\par
The analysis we have presented is rather crude, and the inclusion of showers and detector effects should be considered in order to provide more realistic constraints. The use of NLO matrix elements in the probabilities of eq.~\eqref{eq:probab} could also be taken into account. However, given that the results are particularly encouraging in this simplified approach, we believe that off-shell Higgs production is a very promising probe of the light-quark Yukawa couplings.

At the same time, the sensitivity of the analysis can 
still be improved. For instance in the current off-shell analyses \cite{ CMS:2022ley, ATLAS:2023dnm} also final states with two neutrinos and two leptons are considered, as well as Higgs production in association with jets. The inclusion of more kinematic distributions might be as well helpful. 
\par
Finally, it would be interesting to study off-shell Higgs production in a more global picture from the EFT point of view, hence including more EFT operators in the off-shell Higgs analysis as well as doing a combined fit including various observables that are sensitive to the  light-quark Yukawa couplings.  
\section*{Acknowledgments}
We thank E.~Salvioni and J.~Campbell for useful discussions. We thank L.~Alasfar for providing us the script for fig.~\ref{fig:summary}. R.G.~and M.V.~acknowledge support from a departmental research grant under the project ``Machine Learning approach to Effective Field Theories in Higgs Physics''. The work of
R.G.~has been partially supported by the Italian Ministry of Research (MUR) under contract 2017FMJFMW (PRIN2017). The work of
M.V.~has been partially supported by the MUR  under grant PRIN 20172LNEEZ.
This project has received funding from the European Union’s Horizon Europe research and innovation programme under the Marie Skłodowska-Curie Staff Exchange grant agreement No 101086085 - ASYMMETRY.
\appendix
\section{Results based on $D_s^u$ \label{app:Dsu}}
\begin{figure}[h!]
\centering
     \includegraphics[width= 0.8\textwidth]{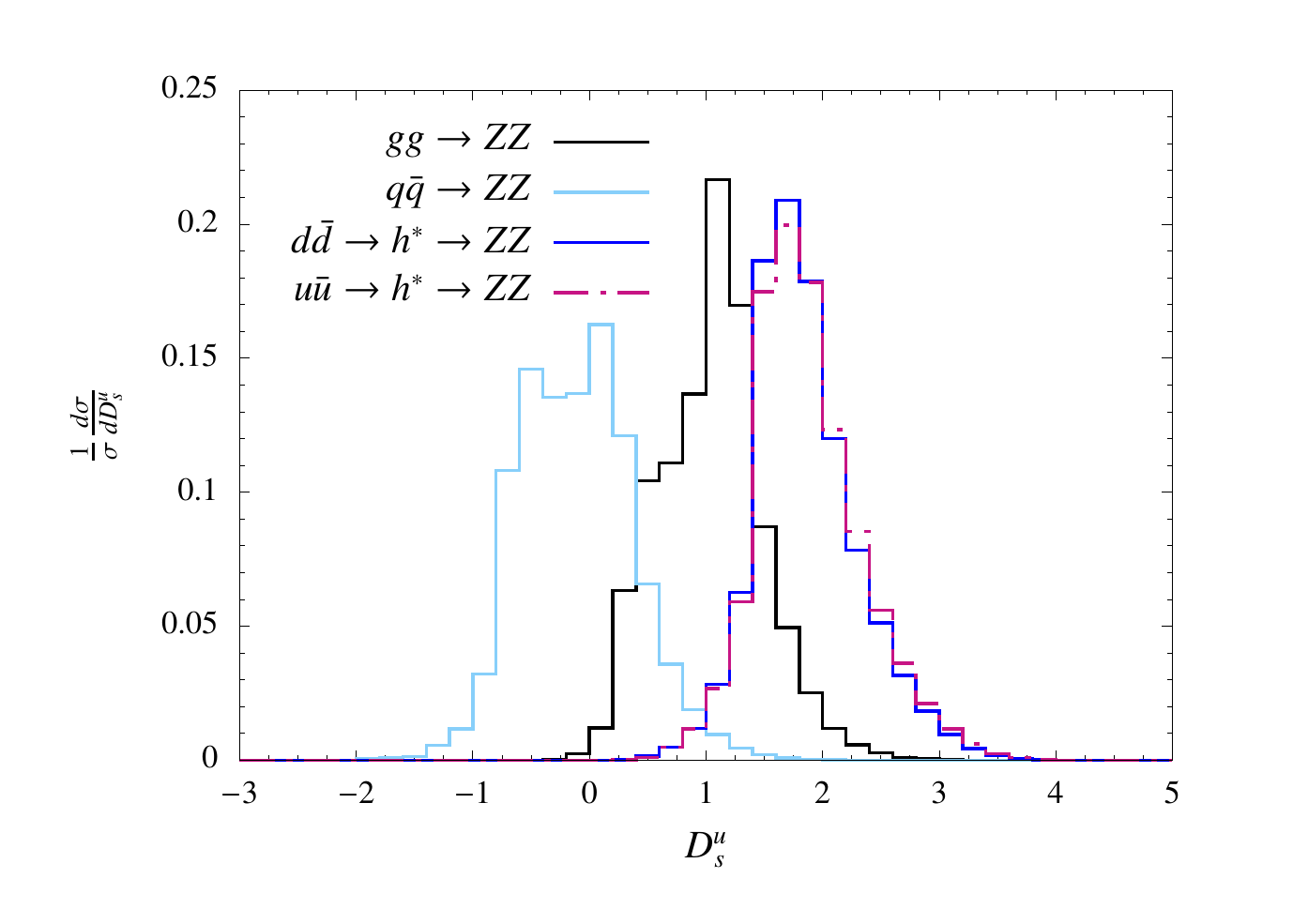}
    \caption{Normalised differential distributions with respect to $D_s^{d}$ for signal (blue and pink dashed) and background (light blue and black) processes. \label{fig:distDSu} }
 \end{figure}

\begin{figure}
\centering
\hspace*{0cm}
 \includegraphics[width=0.5 \textwidth]{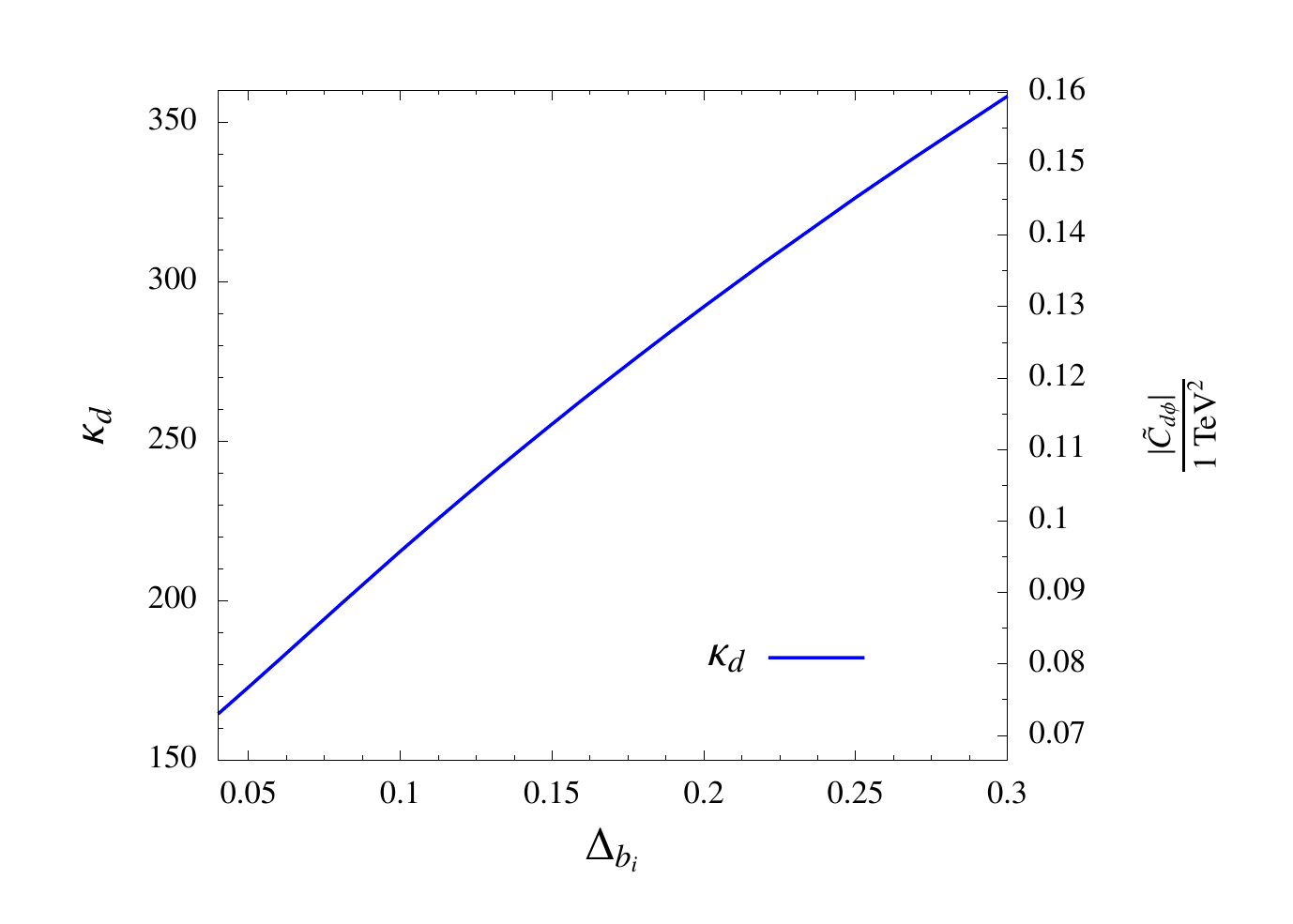} 
\hspace*{-0.5cm}
   \includegraphics[width=0.5 \textwidth]{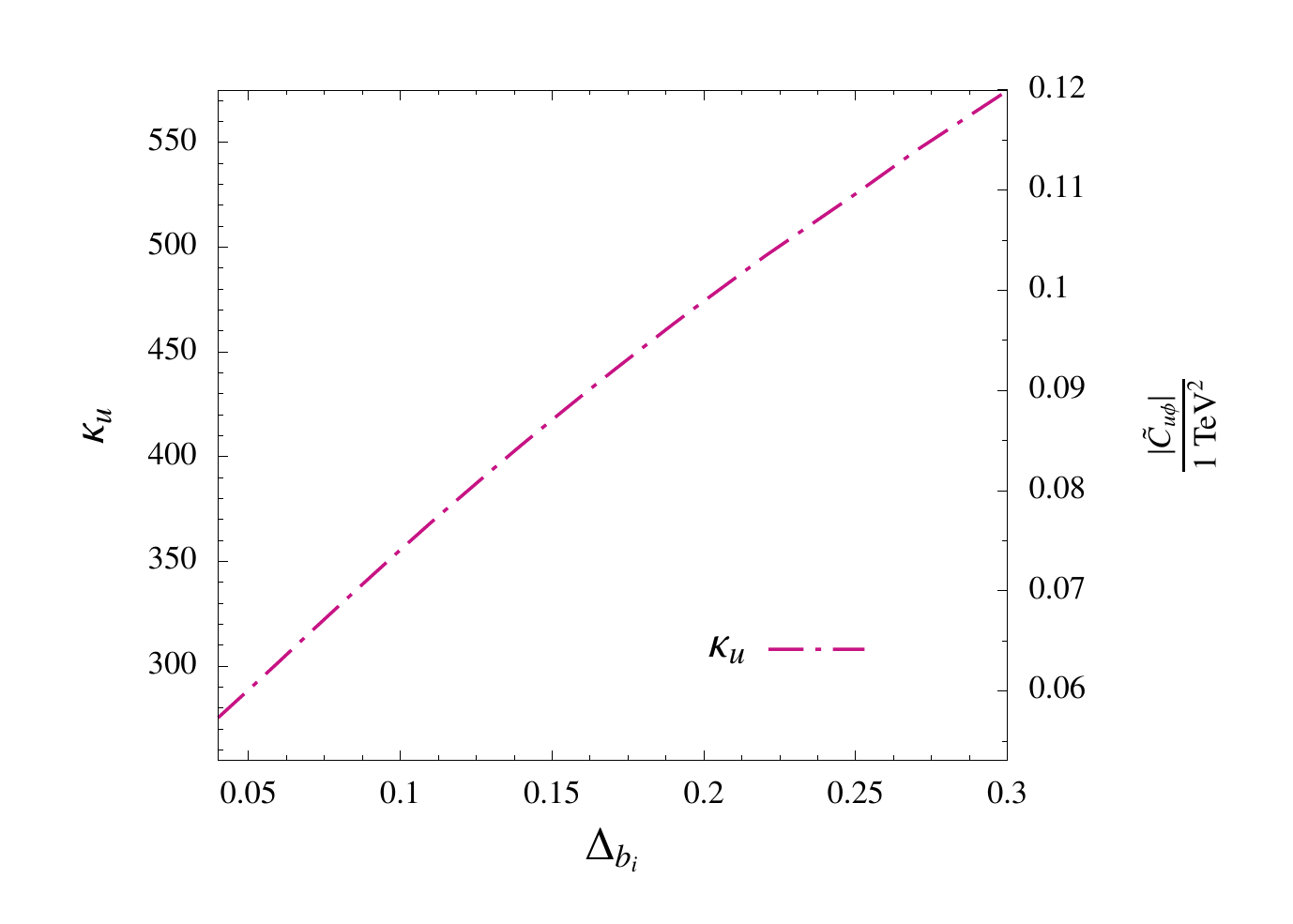} \hspace*{-0.5cm}
 \caption{Dependence of the sensitivity bounds on $\kappa_d$ ($\tilde{C}_{d\phi}$) in the left panel and $\kappa_u$ ($\tilde{C}_{u\phi}$) in the right panel on the assumption made on the size of $\Delta_{b_i}$ using $D_s^u$.  \label{fig:sigmadepDsu}}
\end{figure}

In this appendix we show the results of our analysis using the discriminant $D_s^u$ defined in eq.~\eqref{eq:defds} instead of $D_s^d$. In fig.~\ref{fig:distDSu} we show the distribution of signal and background in $D_s^u$. As for $D_s^d$ we find that the signal peaks at larger values than the background, which shows that the $D_s$ variable is able to discriminate among signal and background. One recognises though that when using $D_s^u$ as discriminating variable the overlap between the signal and the $gg$-initiated background is larger than in case of the $D_s^d$ variable. We find, assuming the optimistic scenario of only systematic error, $\Delta_{b_i}=0.04$,  that  we can restrict 
\begin{align*}
|\tilde{C}_{d\phi}|/(1 \text{ TeV})^2< 0.073/\text{TeV}^2 \quad (\kappa_d< 165 ), \\ 
|\tilde{C}_{u\phi}|/(1 \text{ TeV})^2< 0.057/\text{TeV}^2 \quad (\kappa_u<  275) \hphantom{,}
\end{align*}
at the HL-LHC. These results are slightly worse than those obtained using the $D_s^d$ variable.
 In fig.~\ref{fig:sigmadepDsu} we finally show the dependence of the  $\kappa_d$ and $\kappa_u$ sensitivity bounds on $\Delta_{b_i}$ using the $D_s^u$ variable. Again, we observe some worsening of the sensitivity  in comparison with the analysis based on the $D_s^d$ variable.
\bibliographystyle{utphys.bst}
\bibliography{bibliography}

\end{document}